\documentclass[12pt,a4paper]{article}
\usepackage[utf8]{inputenc}
\usepackage[T1]{fontenc}
\usepackage{amsmath}
\usepackage{amsfonts}
\usepackage{amssymb}
\usepackage{fullpage}
\usepackage{bbold}
\usepackage{xcolor}

\usepackage{natbib}
\usepackage[font={small}]{caption}
\usepackage[affil-it]{authblk}

\newcommand\VaR{\mathit{VaR}}

\usepackage[space]{grffile}
\usepackage{caption}
\usepackage{subcaption}
\usepackage{graphicx}
\usepackage{epstopdf} 
\usepackage{algpseudocode}
\usepackage{booktabs}

\DeclareMathOperator\erf{erf}

\author[1,2]{Christoph Aymanns \thanks{Email: \texttt{aymanns@maths.ox.ac.uk}; Corresponding author}}
\author[1,2,3]{J. Doyne Farmer}
\affil[1]{Institute of New Economic Thinking at the Oxford Martin School}
\affil[2]{Mathematical Institute, University of Oxford}
\affil[3]{Santa Fe Institute}

\title{
The dynamics of the leverage cycle}

\date{ This version: \today \\
First version: July 4, 2014}

\begin{document}
\maketitle

\begin{abstract}
We present a simple agent-based model of a financial system composed of leveraged investors such as banks that invest in stocks and manage their risk using a Value-at-Risk constraint, based on historical observations of asset prices. The Value-at-Risk constraint implies that when perceived risk is low, leverage is high and vice versa, a phenomenon that has been dubbed {\it pro-cyclical leverage}.   We show that this leads to endogenous irregular oscillations, in which gradual increases in stock prices and leverage are followed by drastic market collapses, i.e. a {\it leverage cycle}.  This phenomenon is studied using simplified models that give a deeper understanding of the dynamics and the nature of the feedback loops and instabilities underlying the leverage cycle.  We introduce a flexible leverage regulation policy in which it is possible to continuously tune from pro-cyclical to countercyclical leverage.  When the policy is sufficiently countercyclical and bank risk is sufficiently low the endogenous oscillation disappears and prices go to a fixed point.  While there is always a leverage ceiling above which the dynamics are unstable, countercyclical leverage can be used to raise the ceiling. We also study the impact on leverage cycles of direct, temporal control of the bank's riskiness via the bank's required Value-at-Risk quantile. Under such a rule the regulator relaxes the Value-at-Risk quantile following a negative stock price shock and tightens it following a positive shock. While such a policy rule can reduce the amplitude of leverage cycles, its effectiveness is highly dependent on the choice of parameters. Finally, we investigate fixed limits on leverage and show how they can control the leverage cycle.  
\end{abstract}


\vspace{0.2cm}

\section{Introduction}
Borrowing is essential to economic activity, but leverage is inextricably linked to risk.  On a systemic level, the collective leveraging and deleveraging of financial institutions can lead to booms and busts in asset markets. The recent financial crisis is a case in point for the systemic consequences of the use of leverage. 

Wide-spread deleveraging typically occurs when leveraged investors hit a constraint on their leverage. Such a constraint may arise in a number of ways. If the investor is using collateralized loans to fund its investments, it must maintain margin on its collateral. Alternatively, a regulator may impose a risk contingent capital adequacy ratio. Finally, internal risk management considerations may lead the investor to adopt a Value-at-Risk\footnote{In simple terms Value-at-Risk is a measure of how much the bank could lose with a given small probability.} constraint. All of these cases effectively impose a risk contingent leverage constraint.

Now, suppose there is a negative shock in the asset market associated with an increase in volatility, and as a result the leverage constraint tightens and investors are forced to sell part of their assets\footnote{In principle banks can react in two ways to an increase in market risk: they can raise more capital or sell assets. In practice many banks tended to do the latter, see \cite{Adrian2008a}. This is simply because selling assets can be much easier and faster than raising equity, in particular in times of increasing market risk.}. As investors sell into falling markets they cause prices to fall further. This is the start of a simple positive feedback loop in which selling causes a depression in prices which causes further selling. In a similar way positive news about prices can lead to a decline in perceived risk. This leads to increased leverage which leads to further price increases. This dynamics is referred to as a leverage cycle, see e.g. \cite{Geanakoplos2003} and \cite{Gennotte1990}. A similar dynamic has also been studied in \cite{Brunnermeier2008a}, where the authors investigate the destabilizing feedback between funding liquidity and market liquidity. A further discussion on the destabilizing effects of margin can be found in \cite{Gorton2010}.

It is widely believed that an important driver of leverage cycles lies in the risk management of leveraged investors\footnote{A number of additional drivers of leverage cycles are discussed in the literature. In particular short-termism, herding in financial markets and incentive distortions can play a role in the development of the leverage cycle, see for example \cite{Aikman2012}, \cite{deNicolo2012} and \cite{Gennaioli2012}.}, see for example \cite{Adrian2008a}, \cite{Shin2010} and \cite{Danielsson2001}. Further supporting evidence for the impact of risk management on the leverage cycle can be found in \cite{Tasca2012}, \cite{Adrian2014} and \cite{Colla2011}. However, less is known about the impact of the parameters of risk management on the dynamical properties of leverage cycles and how leverage cycles might be controlled. In order to improve our understanding of the anatomy of leverage cycles, we develop a dynamic computational model of leveraged investors (which we will call banks from now on) that invest in an asset market (which for convenience we call the stock market). Banks have a target leverage that depends on the banks' perceived risk and may be capped at a maximum value. We make the exponent of the relationship between perceived risk and target leverage a parameter of the model. In this way, we are able to capture pro-cyclical leverage policies, which correspond to banks having a Value-at-Risk constraint, and counter-cyclical leverage policies in a single model. Banks are boundedly rational and rely on historical data to estimate the risk of their portfolio. Banks then adjust their target leverage based on changes in perceived risk.

Our model differs from existing models of leverage cycles and pro-cyclical leverage in a number of ways. We develop a fully \textit{dynamic} model of endogenous leverage cycles. This differs from \cite{Geanakoplos2003} and \cite{Geanakoplos2010} who show the existence of leverage cycles in a two period general equilibrium model. Our model shows how a leverage cycle can be a sustained endogenous phenomenon where each cycle sows the seeds for the subsequent leverage cycle.  While our model set-up is similar to \cite{Danielsson2004}, we show that the endogenous dynamics induced by leverage management and historical risk estimation are richer than initially thought. In particular, we observe dynamics ranging from stable fixed points to chaos to unstable behaviour.

Banks in our model rely on historical data for the estimation of their portfolio risk, and are in this sense explicitly boundedly rational. This sets us apart from the model developed in \cite{Zigrand2010} who consider endogenous risk as an equilibrium concept in a financial market with rational VaR constrained investors.  The fact that rational investors can correctly anticipate future volatility and correctly estimate portfolio adjustments needed to reach their leverage targets, means that they settle into a fixed point equilibrium.  In contrast, as we show here, boundedly rational investors will under or overshoot, and the inherent instability of the leverage cycle induces oscillations.   Thus while as shown in \cite{Geanakoplos2010}) bounded rationality is not essential for the existence of leverage cycles, the mismatch between actual market risk and perceived portfolio risk as well as the uncertainty about the extent of price impact have important behavioural consequences that increase the severity of the leverage cycle. The fact that pursuing a target leverage policy can be destabilizing in the case of imperfect knowledge about the extent of market impact has been originally pointed out in \cite{Caccioli2014} and underlies much of the analysis done in this paper.

Another important point of comparison is the agent-based model of leveraged investors developed in \cite{Thurner2010} and \cite{Poledna2013}, which is also dynamic and boundedly rational.  They studied leveraged value investors such as hedge funds that are subject to a leverage ceiling imposed by the lender.  In their model the leverage actually used by investors varies dramatically based on investment opportunities and as a consequence the leverage ceiling is only occasionally reached.  In this model the bank is a dummy agent with infinite capital whose only role is to provide credit to funds.  In contrast, in the model developed here the banks are the key strategic agents.  They always use full leverage, adjusting it as needed to match a regulatory target.  Thus the model introduced here studies the underlying mechanism of the financial crisis of 2008, whereas the earlier model of \cite{Thurner2010} is more relevant to circumstances such as the near meltdown of Long Term Capital Management in 1998.

In contrast to most models in the agent-based literature, we also develop and study simple reduced models. Their simplicity makes them amenable to the tools of dynamical systems theory and allows us to fully characterize the stability properties of the financial system depending on the parameters of our model.  This approach improves our understanding of why the system destabilizes when leverage is high and how changing the banks' risk management can make the system more stable.

Given that volatility is persistent in time, reducing leverage when historical volatility is high is rational from the point of view of a solipsistic individual who is unconcerned about possible systemic effects.  Thus a risk manager who does not think about the market impact of her institution, or that of other similar institutions, will naturally pursue a {\it pro-cyclical} leverage policy.  By this we mean that such an investor will lower leverage when historical volatility is high.  As we explicitly show here, from a dynamical systems point of view this is inherently destabilizing.  

A regulator can potentially correct for this systemic risk by imposing a {\it countercyclical} policy, in which leverage is actually increased when historical volatility is high.   We have set up our model such that we can continuously change from pro-cyclical leverage policies to counter-cyclical leverage policies and explicitly test the impact of different policies on systemic risk.  Countercyclical policies are being widely discussed; our model provides a simple way to get insight into their consequences.  We find that making risk control more countercyclical can indeed be effective in decreasing systemic risk, and provide some insight into how this happens.  To our knowledge we provide the first study of the effect of counter-cyclical leverage on the properties of the leverage cycle.

We also study a scenario in which a regulator adopts a rule to control the banks' required Value-at-Risk quantile. Under this rule the regulator will allow banks to increase their leverage in response to negative price shocks and will require them to decrease their leverage following positive price shocks. In this regime the regulator effectively targets a particular level of asset prices and leverage in the financial system. Our model allows a qualitative evaluation of such a policy rule and provides insight into the circumstances under which it can be effective. 

It is worth commenting that while countercyclical policies and systemic risk are discussed in Basel III, the usage of expected shortfall\footnote{
Expected shortfall is the loss above a given quantile of the return distribution (whereas VaR is the quantile itself).}
is also inherently pro-cyclical, for the same reason that VaR is inherently pro-cyclical.  Thus without an active policy to manage expected shortfall countercyclically,  similar results to those found here should apply.    

The analysis of our model yields five main results:
\begin{itemize}
\item Pro-cyclical leverage management as prescribed in Basel II leads to the endogenous generation of recurring bubbles and crashes in stock prices. These dynamics are driven by the banks' historical risk estimation and leverage adjustment. We refer to this phenomenon as the leverage cycle.
\item The amplitude of leverage cycles, i.e. the extent of price crashes, increases as the banks' riskiness increases.   By increasing this sufficiently (and thereby increasing leverage targets) it is always possible to destabilize the system. 
\item Counter-cyclical leverage policies can stabilize the system under certain conditions on the riskiness of banks but do not solve the issue of leverage cycles in general.
\item Temporal control of the bank's riskiness (via the Value-at-Risk quantile) can decrease the amplitude of leverage cycles but its effectiveness is strongly dependent on the parameter choice for the policy rule.
\item Fixed leverage limits can also curb leverage cycles effectively if set to an appropriately low level.
\end{itemize}

Before proceeding, we would like to mention that the financial model presented here can be coupled to a macroeconomic model in order to study the spill-over effects of the leverage cycle from the financial sector to the real economy. In fact, an unpublished earlier version of this paper studies exactly this phenomenon. In that paper, we couple the financial model developed here with a simple macroeconomy by allowing banks to give loans to firms in the real economy. Via this credit channel the financial leverage cycle affects the activity in the real economy. We observe that, provided that the banks' stock and loan portfolios are sufficiently linked, leverage cycles originating in the stock market spill over into the real economy causing cycles in credit provision and output. We have not included these results here because they are not essential for the main results and the macromodel complicates the exposition.

The remainder of this paper is organized as follows. In section \ref{MODEL} we outline the full multi-asset, agent-based model of the financial system. In section \ref{lev_cycles_m} we will study the dynamics of the full agent-based model. We will then develop two reduced models in section \ref{Drivers} and study the parameter dependence of the system dynamics, and conclude in section 5.

\section{Multi-asset model of a financial system}\label{MODEL}
We propose a simple model of a financial system consisting of leveraged investors and a noise trader. In the following we will refer to these investors as banks. Banks are fundamentalists and distribute their assets across a range of tradable securities that provide an exogenous, stochastic dividend stream. For simplicity we refer to these securities as stocks. We focus our analysis on the dynamics induced by the leverage management and portfolio allocation of the banks. Therefore, we abstract from the influence other financial institutions may have on the market dynamics and subsume their activity in the noise trader. 

Crucially, banks are boundedly rational and have imperfect information about the stock market. Therefore, banks must learn about the stock market by analysing past observations of market behaviour. In particular, in order to manage their risk, banks estimate the covariance matrix of their portfolio from historical price movements. Banks then adjust their leverage based on their perception of portfolio risk. Due to the historical estimation banks' perceived portfolio risk may be drastically different from their actual risk. As we will demonstrate below, this mismatch between perception and reality of market conditions plays an important role in the dynamics of the model. In the following sections we will outline the behaviour of the banks and the noise trader more formally.

\subsection{Bank}
\subsubsection{Accounting}
The financial system is comprised of a set of banks indexed by $j \in \left\lbrace 0,..., N_b \right\rbrace$. For ease of exposition we will drop this index wherever possible. Banks are characterized by their balance sheet, their investment strategy, and potential regulatory requirements. Banks hold a portfolio of two types of assets: cash and shares in stocks. Cash is a non-interest bearing risk-free asset, while stocks are risky assets. More formally, the asset side of the bank balance sheet is given by
\begin{eqnarray*}
\mathcal{A}_{t} &=& c_{t} + \mathbf{n}_{t}^T \mathbf{p}_t, \\
\mathbf{n}_{t} &=& \left(n_{1,t},...,n_{N_f,t} \right)^T,\\
\mathbf{p}_{t} &=& \left(p_{1,t},...,p_{N_f,t} \right)^T,
\end{eqnarray*}
where $c_{t}$ is the cash investment, $\mathbf{n}_{t}$ the vector of stock ownership, $\mathbf{p}_t$ is vector of share prices, $N_f$ is the number stocks. Note that the total number of shares of a given stock is normalized to $1$ such that
\begin{equation*}
\sum_{j=1}^{N_b} n_{ji,t} = 1 \hspace{0.5cm}\forall i,t.
\end{equation*}
Also note that banks face a long only constraint, i.e. $n_{ji,t} \geq 0$. This implies that banks cannot short an asset which they consider overpriced. This limits the extent of arbitrage that is possible in this setting. Mispricing may therefore persist longer than in the case where short selling is permitted. By definition the equity of the bank is given by
\begin{equation*}
\mathcal{E}_{t} = \mathcal{A}_{t} -\mathcal{L}_{t},
\end{equation*}
where $\mathcal{L}_{t}$ generically represents all liabilities of the bank at time $t$. These liabilities could be composed of deposits, interbank loans or other forms of short term debt. 
We assume for simplicity that the banks do not face funding restrictions, i.e. should a bank decide to increase its liabilities it always finds willing lenders. This is of course a strong simplifying assumption. If the bank faces exogenous constraints on its funding its ability to increase its leverage will be impaired. In this model, our focus is on the impact of risk management on the stability of the system. Therefore we explicitly try to exclude exogenous constraints if possible. By assuming no funding restrictions we effectively establish a lower bound on the system's stability. 

Finally, we define the bank's leverage as follows:
\begin{equation*}
\lambda_{t} = \frac{\mathcal{A}_{t}-c_{t}}{\mathcal{E}_{t}}.
\end{equation*}

Following \cite{LeBaron2012} and \cite{Hommes}, the budget constraint of the bank is given by
\begin{eqnarray*}
\mathcal{E}_{t} &=& c_{t} + \mathbf{n}_{t}^T \mathbf{p}_t  - \mathcal{L}_{t} \\
&=& c_{t-1} + \mathbf{n}_{t-1}^T \left( \mathbf{p}_t + \boldsymbol{\pi}_{t-1} \right) - \mathcal{L}_{t-1}(1+r_{\mathcal{L},t-1}) - d_{t-1},
\end{eqnarray*}
where $\boldsymbol{\pi}_{t-1}$ is the $N_f \times 1$ vector of stock dividends. $r_{\mathcal{L},t-1}$ is a generic interest rate that the bank has to pay on its debts. Finally $d_{t-1}$ represents a dividend paid to shareholders. We refer to the expression $\mathcal{L}_{t-1}r_{\mathcal{L},t-1} + d_{t-1}$ as the banks' funding cost.

For convenience we assume that each bank's income through dividends is equal to its funding cost, i.e.: $\mathbf{n}_{t-1}^T \boldsymbol{\pi}_{t-1} = \mathcal{L}_{t-1}r_{\mathcal{L},t-1} + d_{t-1}$. This amounts to assuming that the banks immediately pay all their income earned through dividends paid as part of stock ownership as dividends and interest to their shareholders and lenders. However, this excludes income through valuation gains from stock trading. This is a strong assumption but simplifies the decision problem of the bank significantly and allows for variable bank equity while ignoring the market for bank debt for now. In section \ref{2D} we will simplify this assumption further and assume that the equity of banks is fixed.

This assumption implies that equity only changes via changes in the prices of the stocks in which the bank holds long positions. In particular, this assumption yields the following simplified budget constraint:
\begin{equation}
\label{budgetConst}
\begin{aligned}
\mathcal{E}_{t} &= c_{t} + \mathbf{n}_{t}^T \mathbf{p}_t - \mathcal{L}_{t}  \\
&= c_{t-1} + \mathbf{n}_{t-1}^T \mathbf{p}_t - \mathcal{L}_{t-1}.
\end{aligned}
\end{equation}

\subsubsection{Expectation formation}
In order to form expectations about future conditions of the stock market, banks rely entirely on the analysis of historical market data. In particular, banks use observations of past dividends and prices to compute the expected dividend price ratio, stock price return and variance of stock prices.

Banks are fundamentalist investors and as such base their investment decision on the expected dividend price ratio of a stock. The expected dividend price ratio is computed as an exponential moving average of past dividend price ratios. In particular we have for the dividend price ratio of stock $i$:
\begin{equation}
\hat{r}_{i,t+1} = (1 - \gamma) \hat{r}_{i,t} + \gamma \frac{\Pi_{i,t}}{p_{i,t}},
\end{equation}
where $\Pi_{i,t}$ is the dividend paid by stock $i$ at time $t$ and $p_{i,t}$ is the corresponding stock price. We typically choose $\gamma \sim 0.1$.

In a similar fashion, banks estimate the covariance matrix of the stocks price returns. Banks rely on the covariance matrix of stock prices for its risk management as will be discussed in section \ref{leverageDynamics}. The method used here is comparable to the RiskMetrics approach described in \cite{Morgan1996} and \cite{Andersen2006}. Consider the log return of the price of stock $i$: $x_{i,t} = \log\left(\frac{p_{i,t}}{p_{i,t-1}} \right)$ which are the components of the vector $\mathbf{x}_t$. First, we estimate the conditional sample mean of the return vector by
\begin{equation}
\label{expR}
\hat{\boldsymbol{\mu}}_t = \delta \mathbf{x}_{t-1} + (1- \delta) \hat{\boldsymbol{\mu}}_{t-1},
\end{equation}
where $\delta < 1$ determines the horizon of this exponential moving average. In practice the conditional sample mean of the stock price return is almost always zero. We estimate it anyway in order ensure unbiased estimates of the covariance matrix in all cases. We now estimate the conditional sample covariance matrix of returns by
\begin{equation}\label{EQ::Cov_est}
\mathbf{\Sigma}_t = \delta \left(\mathbf{x}_{t-1} - \hat{\boldsymbol{\mu}}_{t} \right) \left(\mathbf{x}_{t-1} - \hat{\boldsymbol{\mu}}_{t} \right)^T +(1-\delta) \mathbf{\Sigma}_{t-1}.
\end{equation}
The bank then simply assumes that the covariance matrix in the next time step is the same as the current estimate of the covariance matrix, i.e.:
\begin{equation}
\mathbf{\Sigma}_{t+1} \approx \mathbf{\Sigma}_{t}.
\end{equation}
Note the distinction between the two concepts of returns covered in this section. On the one hand the bank uses price dividend ratios for its investment allocation. On the other hand it relies on stock price returns for its risk management as it has to protect itself against stock price devaluations.

\subsubsection{Portfolio choice}
\label{portDec}
Banks use a heuristic rule to compute their portfolio choice as a function of the expected risk-return relationship of the stocks. Banks compute the risk-return ratio as follows:
\begin{equation}
s_{i,t+1} = \hat{r}_{i,t+1} / \sigma_{i,t+1},
\end{equation}
where $\sigma_{i,t+1}^2 = \Sigma_{ii,t+1}$ is the $i$th diagonal element of the covariance matrix and $\hat{r}_{i,t+1}$ is the expected dividend price ratio. Hence the bank's decision criterion is simply the Sharpe ratio of the stock. The portfolio weight $w_{i,t}$ of stock $i$ is then given by the following rule:
\begin{equation}
w_{i,t} = (1 - w_c) \frac{\exp(\beta s_{i,t+1})}{\sum_j \exp(\beta s_{j,t+1})},
\end{equation}
where $w_c$ is a fixed weight for the bank's cash reserve. In an alternative specification the bank's portfolio choice is determined by numerically optimizing a mean variance portfolio with ``no short sale'' constraints\footnote{Note that the mean variance portfolio problem cannot be solved by the usual Lagrangian approach in the presence of ``no short sale'' constraints as required here. Therefore a numerical optimization method has to be used to determine the efficient portfolio.}. We found that the key results of this paper do not depend on the specific rule for the portfolio choice. Therefore we chose the more computationally efficient rule described above. In Appendix \ref{AppendixPortfolio} we describe the alternative, optimizing portfolio choice specification and show that the results remain unchanged.

\subsubsection{Risk management}
\label{leverageDynamics}
Risk management is an important component of any financial institution's activities. Here, we assume that banks use a Value-at-Risk (VaR) approach to control their exposure to the stock market. In particular, banks will try to ensure that their equity exceeds their Value-at-Risk in order to protect themselves against large losses on the stock market. This behaviour may be due to a regulatory constraint as proposed in \cite{Zigrand2010}, \cite{Adrian2012} and \cite{Corsi2013}. However, even in the absence of regulation, financial institutions are likely to control their Value-at-Risk. Rather than determining the existence of a VaR type risk management approach, regulation can then be thought of putting constraints on the parameters of the VaR model.

For simplicity, we begin by assuming that banks believe that stock returns are normally distributed with zero mean\footnote{Note that in practice we do find that the conditional sample mean of the stock price return is almost always zero making the zero mean assumption model consistent.}. In this simple case we have for the per-dollar VaR
\begin{equation}
\VaR_{a,t} = \sqrt{2} \hat{\sigma}_{P,t} \erf^{-1}(2a-1),
\end{equation}
where $a$ is the VaR quantile and $\sigma_{P,t}$ is the per-dollar conditional estimate of the portfolio standard deviation at time $t$. $\sigma_{P,t}$ is determined as follows:
\begin{equation}
\sigma_{t,P}^2 = \mathbf{w}_t^T  \mathbf{\Sigma}_t  \mathbf{w}_t,
\end{equation}
where $\mathbf{w}_t = (w_{1,t},...,w_{N_f,t})^T$. The bank tries to ensure that its VaR does not exceed its equity, hence:
\begin{equation}
\VaR_{a,t} \mathcal{A}_t \leq \mathcal{E}_t.
\end{equation}
For a bank that maximizes its return on equity this constraint will be binding. The bank then has the following target leverage:
\begin{equation}\label{EQ::lev_t_1}
\overline{\lambda_t} = \VaR_{a,t}^{-1} = \left( \sqrt{2} \sigma_{P,t} \erf^{-1}(2a-1) \right)^{-1} = \alpha / \sigma_{P,t},
\end{equation}
where we replaced the constant $( \sqrt{2} \erf^{-1}(2a-1))^{-1}$ by $\alpha$. From equation \ref{EQ::lev_t_1} we see that the bank's target leverage is inversely proportional to its perceived portfolio variance. 

While we derived equation \ref{EQ::lev_t_1} assuming normally distributed Gaussian returns, it is more general and includes both the Gaussian case as well as the maximally heavy-tailed symmetric return distribution with finite variance. In the Gaussian case for a confidence interval $a = 0.99$ we obtain $\alpha \approx 0.42$. Now consider the maximally heavy tailed return distribution. According to Chebychev's inequality we have for the probability that the loss $X$ exceeds $k \sigma$:
\begin{equation}
P( X > k \sigma) \leq \frac{1}{2k^2}.
\end{equation}
In the maximally heavy tailed case the upper bound is tight. For a confidence interval of $a=0.99$ we have $P( X > k \sigma) = 0.01$. Thus $k \approx 7.07$ and $\alpha \approx 0.14$. Therefore from the perspective of the bank, changing its beliefs about how heavy-tailed the return distribution is, does not fundamentally change the relationship between the perceived risk and leverage. Instead it only scales down leverage; the bank effectively becomes a more cautious investor. 

In a given time step the bank evaluates the difference to the desired target size of its balance sheet: $\Delta \mathcal{B}_t = \overline{\lambda_t} \mathcal{E}_t - \mathcal{A}_t$. If $\Delta \mathcal{B}_t > 0$ we assume that the bank simply raises the necessary funds in order to increase its assets and liabilities by the desired amount $\Delta \mathcal{B}_t$. If $\Delta \mathcal{B}_t < 0$ the bank will sell part of its assets in the next round in order to reduce its assets and liabilities by the desired amount.

\subsubsection{Generalized target leverage}
In the following, we generalize the expression for the target leverage in Eq. \ref{EQ::lev_t_1} by introducing three potential policy parameters. (1) We add a constant $\sigma_0$ to the expression for the perceived risk, (2) we generalize the exponent of the perceived risk by setting it to the parameter $b$ and (3) we make $\alpha$ a function of time such that $\alpha \rightarrow \alpha_t$. We refer to parameter $b$ as the cyclicality parameter\footnote{Note that we do not claim that these changes can be readily derived from risk management first principles. Instead we consider a scenario in which a regulator has control over the parameters of the functional form derived based on traditional risk management. Assuming the regulator has control over these parameters, the question is then how these parameters should be chosen or controlled over time in order to minimize leverage cycles.}. After these steps we obtain the generalized expression for the target leverage:
\begin{equation}
\label{EQ::lev_target}
\overline{\lambda_t} =  \alpha_t \left( \sigma_{P,t}^2 + \sigma_0 \right)^b.
\end{equation}
$\alpha_t$ determines the scale of the bank's leverage. We can thus think of it as a proxy for the bank's riskiness and refer to $\alpha_t$ as the bank's risk parameter. Throughout the majority of this paper we will take $\alpha$ pre-determined, i.e. $\alpha_t = \alpha$. In section \ref{temp_con} we relax this assumption and study how $\alpha_t$ can be controlled to mitigate leverage cycles. In the simple case of normally distributed returns and fixed $\alpha$ we have $\alpha = ( \sqrt{2}\erf^{-1}(2a-1))^{-1}$.

\begin{figure}
\centering
\includegraphics[width=0.5\textwidth]{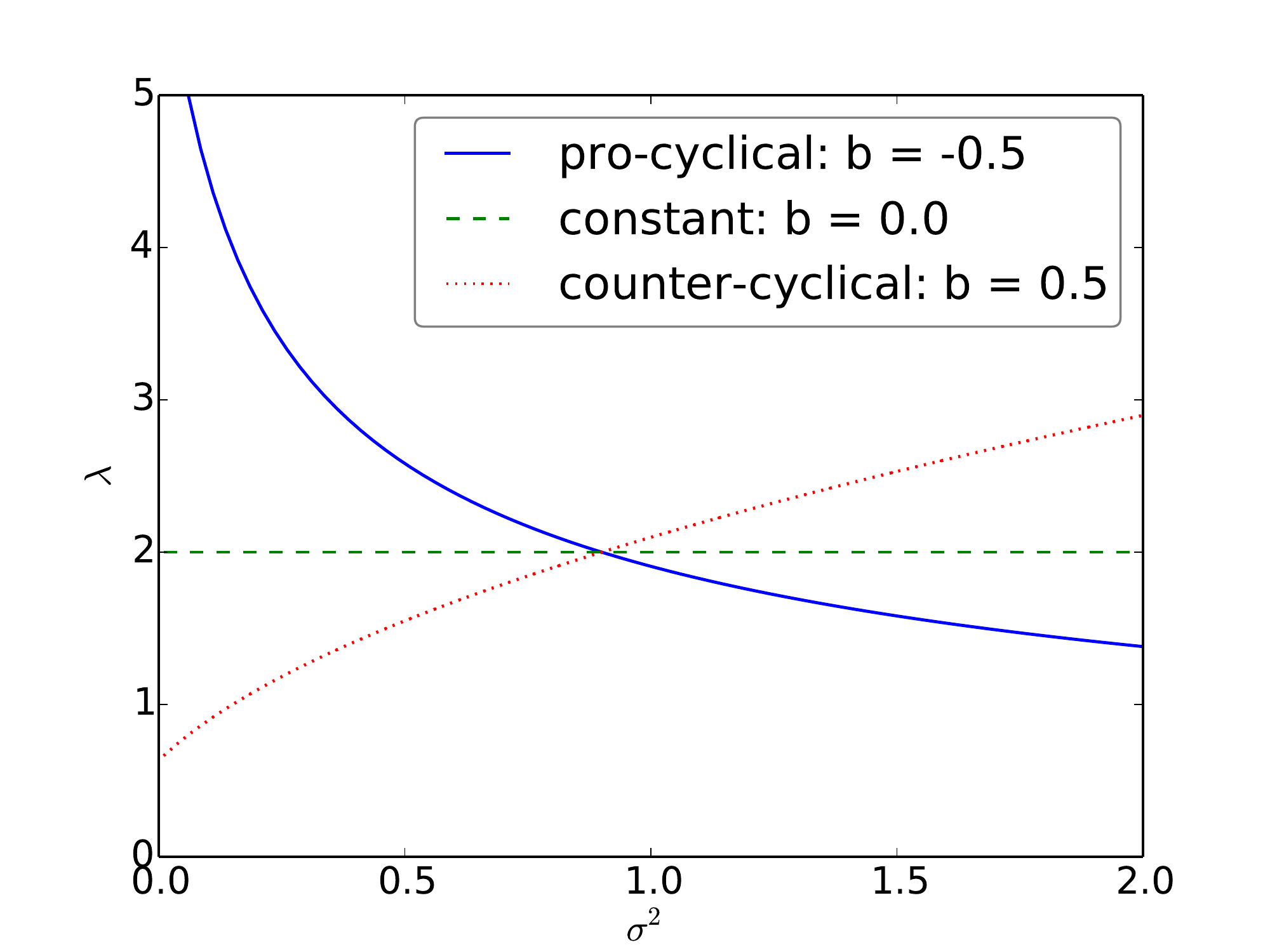}
\caption{Relationship between perceived risk (x-axis) and target leverage (y-axis) as given in equation \ref{EQ::lev_target} for different values of $b$. For all three cases we take $\sigma_0 = 0.1$. We refer to the case where $b=-0.5$ as the pro-cyclical leverage case. $b=0$ is the constant leverage case and $b=0.5$ is the counter-cyclical case.}
\label{FIG:intuitionbs_0}
\end{figure}

As can be seen in figure \ref{FIG:intuitionbs_0} the cyclicality parameter $b$ determines how perceived portfolio risk is related to the target leverage. In the case of Gaussian returns we have $b = -0.5$. For negative $b$ leverage will decrease with perceived portfolio risk. We refer to this situation as the pro-cyclical case. Conversely, for positive $b$ leverage will increase with perceived portfolio risk. We call this the counter-cyclical case. Finally, $\sigma_0$ is a lower bound on the perceived portfolio risk. In the pro-cyclical case this is equivalent to a cap on leverage, while in the counter-cyclical case it corresponds to a minimum leverage. In most cases we take $\sigma_0 = 0$. Later on we interpret $\sigma_0$ as a policy parameter and investigate its impact on the dynamics of the stock market.

\subsection{Noise trader}\label{SEC::noise_trader}
In order to control the market power of the banks, we introduce a noise trader. The noise trader can be thought of representing a population of hedge or mutual funds in the market with fundamentalist trading patterns that are not fully modelled. This partial modelling introduces a random component in its portfolio allocation. Hence we refer to it as a noise trader. In this model the noise trader effectively acts as market maker for the stock market.

Similar to banks the assets of the noise trader consist of a stock portfolio and a cash reserve: $A_{N,t} = c_t + \mathbf{n}_{t}^T \mathbf{p}_t$. The noise trader's stock portfolio weights $w_{i,t}$ are computed as follows:
\begin{equation}
w_{i,t+1} = (1-w_c) \frac{v_{i,t+1}}{ \sum_j v_{j,t+1}} ,
\end{equation}
where $v_{i,t}$ evolves according to a stochastic process and is not necessarily bounded between zero and one. Therefore we normalize $v_{i,t}$ and multiply by one minus the required cash weight $w_c$ to obtain the portfolio weight for stock $i$. Specifically, $v_{i,t}$ evolves according to the following process:
\begin{equation}
\begin{aligned}
\frac{dv_{i,t}}{v_{i,t}} &=  \underbrace{ \rho (1 / N_f - v_{i,t})}_{\text{Portfolio balance}} +  \underbrace{ \zeta ( \hat{r}_{i,t+1} - \sum_j \hat{r}_{j,t+1} / N_f ) }_{\text{Fundamentalist}} + \underbrace{\eta dW}_{\text{Noise}}, \\
v_{i,t+1} &= v_{i,t} + dv_{i,t}.
\end{aligned}
\end{equation}
$\rho < 1$ determines how quickly the portfolio returns to a balanced portfolio. $\zeta$ is a scaling parameter to ensure that  contribution of the fundamentalist term is of comparable size to the other terms in the equation. $\hat{r}_{i,t+1}$ is the expected return and is computed in the same ways as for banks. Finally $dW \sim \mathcal{N}(0,1)$ is a standard Brownian motion term while $\eta$ determines the standard deviation of the random walk.

This specification of the noise trader's portfolio weights ensures that the weights do not diverge (portfolio balance term) and are economically sensible (fundamentalist term). In particular the second term ensures that the noise trader is a weak fundamentalist investing a larger fraction of its portfolio into stocks with higher than average return.

Due to this fundamentalist component and due the fact that the noise trader is unleveraged, it stabilizes the stock market. First consider the impact of the fundamentalist component of the weight update: If the price of a stock rises significantly, its dividend price ratio will decrease relative to other stocks. The noise trader will then shift its portfolio away from this stock. This process leads to a mean reversion in stock prices. The fact that the noise trader is not leveraged means that it acts as a counter-weight to the bank which will actively manage its leverage and thereby destabilise the market as we will show in section \ref{lev_cycles_m}.

\subsection{Dividend process}
The dividend process follows a simple geometric random walk with drift
\begin{equation}
\frac{d \pi_{i,t}}{\pi_{i,t}} = \mu + \varphi dW,
\end{equation}
where the drift term $\mu \approx 0$ and the standard deviation of the Brownian motion term $\varphi \ll 1$.

\subsection{Stock market mechanism}
\label{marketMech}
In the following we propose a simple linear multi-asset market clearing mechanism that is based on the single asset markets developed in \cite{Hommes} and \cite{LeBaron2012}. The linearity of the approach allows efficient computation of prices and makes this approach scalable to potentially many assets. 

First consider a single stock. As mentioned earlier the total supply of shares of a given stock is normalised to $1$. In monetary units the supply of stocks, i.e. the number of shares in the market times their price, is therefore simply given by the market clearing price. The demand, in monetary units, for stock $i$ is the sum over all banks' portfolio allocations to this stock, i.e.
\begin{equation}\label{demand}
\mbox{Demand}_{i,t+t} = \sum_{j \in \mathcal{I}} w_{ji,t+t} \mathcal{A}_{j,t+t},
\end{equation}
where $\mathcal{I}$ is the set of all investors in the stock market, i.e. in this case banks and the noise trader. $w_{ji,t+1}$ is investor $j$'s portfolio weight for stock $i$. Finally $\mathcal{A}_{j,t+1}$ is the value of investor $j$'s assets after market clearing.

In order to compute the market clearing price for stock $i$ we equate demand (eq. \ref{demand}) and supply ($p_{i,t+1}$) in monetary units. We substitute into equation \ref{demand} the value for the investors' assets $\mathcal{A}_{j,t+t}$ taking into account any changes in assets due to additional leveraging or deleveraging. We then obtain:
\begin{equation}\label{EQ::stock_price}
p_{i,t+1} = \sum_{j \in \mathcal{I}} w_{ji,t+1} \left( c_{j,t} + \mathbf{n}_{j,t}^T \mathbf{p}_{t+1} +  \Delta\mathcal{B}_{j,t} \right).
\end{equation}
Note that the banks' demand depends on the price of the stock as it determines the wealth of the bank. As discussed above, we assume that the investor chooses the portfolio weights prior to market clearing, i.e. the weights are not a function of the price computed in market clearing. Furthermore, banks compute their desired asset change prior to market clearing. Economically this corresponds to the case in which banks have to commit to their actions prior to receiving information about the market price. In this sense banks have bounded rationality during the market clearing process. This approximation linearises the problem as it removes the price dependence of the bank's decision parameters. Expressed in matrix form we have:
\begin{equation}
\mathbf{p}_{t+1} = \mathbf{W}_{t+1} \left(\mathbf{N}_t \mathbf{p}_{t+1} + \mathbf{c}_{t} + \Delta\mathcal{B}_{t}\right),
\end{equation}
where $\mathbf{W}$ is the $N_f \times |\mathcal{I}|$ matrix of stock portfolio weights, $\mathbf{N}$ is the $|\mathcal{I}| \times N_f$ stock ownership matrix, $\mathbf{c}$ is the vector of cash reserves and $\Delta\mathcal{B}_{t}$ is the vector of asset changes. Thus the market clearing vector of stock prices is given by:
\begin{equation}
\label{marketClearingPrice}
\begin{aligned}
\mathbf{p}_{t+1} &= \mathbf{U}_{t+1}^{-1} \mathbf{W}_{t+1} ( \mathbf{c}_t  + \Delta\mathcal{B}_{t}) \\
\mathbf{U}_{t+1} &= \mathbb{1} - \mathbf{W}_{t+1} \mathbf{N}_t
\end{aligned}
\end{equation}
Note that for the stock prices to be well defined two conditions have to be satisfied: (1) There must exist a bank $j$ such that $c_{j,t} > 0$ and (2) the matrix $\mathbf{U}$ must be invertible. In practice these conditions are always satisfied. 

\section{Leverage cycles in the multi-asset model}\label{lev_cycles_m}
In this section we will study the dynamics of the multi-asset model with one bank, three stocks and a noise trader. Below we will briefly comment on the choice of parameters for this section. All parameters are summarized in table \ref{TAB::param_overview}. We will then proceed to discuss the results of the simulations.

\subsection{Simulation set up}\label{sim_Set_up}
\begin{table}
\begin{tabular}{lllr} 
\toprule
Agent & Parameter \\ 
\cmidrule{2-4}
& Notation & Description & Value \\
\midrule
Bank & $\delta$ & Memory parameter in covariance estimation & $0.1$ \\ 
	 & $\alpha$ & risk parameter & $0.1$ \\
	 & $b$ & cyclicality parameters & $-0.5$ \\
	 & $\mathcal{E}_{0}$ & Initial bank equity & $23$ \\ 
	 & $\lambda_{0}$ & Initial bank leverage & $5$ \\ 
	 & $w_c$ & Cash weight & $0.2$ \\ 
	 & $\gamma$ & Memory parameter in return estimation & $0.1$ \\ 
	 & $\beta$ & Intensity of choice in portfolio allocation & $0.1$ \\
	 & $N_b$ & Number of banks & $1$ \\ 
\midrule
Noise trader & $A_{N,0}$ & Initial noise trader assets & $115$ \\ 		
	 		 & $\rho$ & Portfolio balance parameter & $0.05$ \\ 	
			 & $\zeta$ & Fundamentalist parameter & $5$ \\ 	
			 & $\eta$ & Standard deviation Brownian motion & $0.2$ \\ 
\midrule
Dividend process & $\mu$ & Dividend drift term & $10^{-5}$ \\ 		
	 		 & $\phi$ & Standard deviation & $0.05$ \\ 	
	 		 & $N_S$ & Number of stocks & $3$ \\ 	
\bottomrule 
\end{tabular} 
\caption{Overview of simulation parameters for full model. The most important parameters for dynamics of the model are $\delta$, $\alpha$ and $b$. The initial conditions listed are only relevant in setting the relative sizes of the bank to the noise trader. This is important in order to stabilize the dynamics of the model.}
\label{TAB::param_overview}
\end{table}

\paragraph{Banks:} We run the simulation with only one bank since we are not explicitly studying the impact of heterogeneity among banks. In the homogeneous case we are considering here several banks could always be collected into one representative bank. The initial equity of the bank $\mathcal{E}_0$ sets the monetary scale of the simulation and has no impact on the actual dynamics of the simulation.  We therefore just pick an arbitrary number. We choose an initial leverage $\lambda_0$ that lies roughly in the order of magnitude of a real bank. The initial assets are then determined by the initial leverage and equity. We calibrate $\gamma$ to introduce some momentum into the bank's portfolio weights. 

We choose $\alpha$ to obtain roughly realistic levels of leverage in the simulation. Finally the value for the memory parameter in the exponential moving average to estimate the portfolio covariance matrix corresponds roughly to the value recommended in the original manual on the RiskMetrics approach, see \cite{Morgan1996}.

\paragraph{Noise trader:} We choose the size of the noise trader such that its market power is sufficient to stabilize the market dynamics somewhat while not dominating them. The remaining parameters are calibrated such that the noise trader on average neither loses nor gains money. In general we find that the investor with more stable portfolio weights tends to accumulate money over the course of the simulation. In order to counteract this tendency we increase the standard deviation $\eta$ in the Brownian motion term of the weight update rule until we find that on average neither agent accumulates equity.

\subsection{Leverage cycles}

In order to demonstrate how leverage management can affect the model dynamics, we will contrast two cases:
\begin{enumerate}
\item Passive case: The bank does not manage its leverage but invests into stocks according to its fundamentalist portfolio rule. This means that the bank's leverage changes passively as stock prices changes. If prices go up, the bank's leverage will go down and vice versa. In particular, a bank will not borrow more if its leverage goes down, nor will it sell assets when its leverage goes up. 
\item Active case: The bank actively manages its leverage as outlined in section \ref{leverageDynamics} and invests into stocks according to their fundamentalist portfolio rule. This means that the bank has a target leverage that is determined by its perceived portfolio risk. If it is above its target leverage it will sell assets and repay part of its debt while it will borrow more and invest more if it is below its target leverage.
\end{enumerate}

We wish to emphasize that the passive case is manifestly unrealistic, and is merely a reference point for comparison.  Over the long run the leverage of such a bank would make a random walk, and after a sufficiently long time would eventually become so large to cause the bank to default.  Nonetheless, this provides an interesting point of comparison because such a bank does not have to do any trading to adjust its leverage, and therefore has no systemic impact.  Thus in a sense this represents a ``pure" case.  Since we are only simulating for a short period there is not enough time for the leverage to eventually reach large values.

In the following we will document the qualitative behaviour of the model by studying exemplary time series. In the subsequent sections we will study the model in more detail, develop an intuition on the drivers of the dynamics induced by leverage management and study the parameter dependence of the dynamics.

\begin{figure}
\centering
\includegraphics[width=0.5\textwidth]{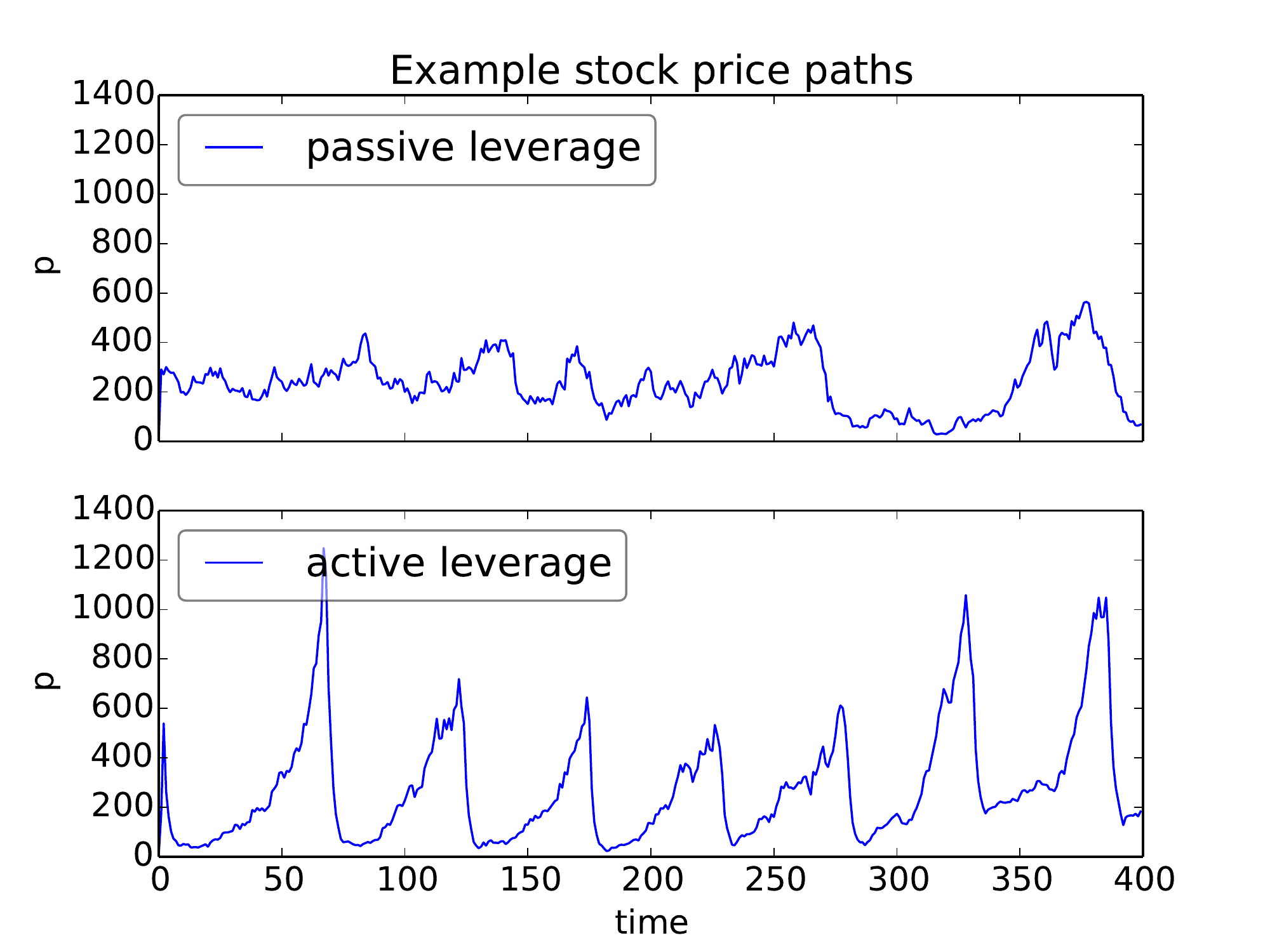}
\caption{Comparison of two exemplary time series for the price of an individual stock. Top: Bank does not manage its leverage. Bottom: Bank actively manages its leverage. Clearly the variation in the active case is much larger than in the passive case. When the bank manages its leverage prices undergo recurring patterns of relatively gradual increase in price followed by drastic price crashes. When the bank does not actively manage its leverage, prices are driven exclusively by the portfolio adjustment of the bank to changing dividend price ratios and the activity of the noise trader.}
\label{FIG:active_passive}
\end{figure}

In figure \ref{FIG:active_passive} we compare two time series for the price of an individual stock generated by the model starting from the same initial conditions. The top panel shows case 1 in which the bank does not manage its leverage. This means that the bank does not have a target leverage and its leverage simply changes as the value of its assets changes: if the value of the assets appreciates the leverage of the bank decreases and vice versa. In contrast to the bank that actively manages its leverage, we have for the passive bank's change in assets through leveraging or deleveraging $\Delta \mathcal{B}_t = 0$ throughout. Therefore, in the top panel the price movements are solely driven by the bank and noise traders portfolio adjustment.

The bottom panel corresponds to case 1 in which the bank actively manages its leverage. Recall that under active leverage management the bank reacts to decreases in perceived risk by increasing its target leverage. Similarly, it decreases its target leverage when perceived risk increases. Furthermore, the bank then adjusts the size of its asset portfolio in order to reach its desired target leverage by an amount $\Delta \mathcal{B}_t$.  

By simple visual comparison the difference in the two price time series is apparent. In the case of active leverage management the price dynamics are characterised by recurring gradual increases in price followed by rapid and drastic collapses in price. The price dynamics in the passive case are driven by bank's portfolio choice, the underlying driving dividend process and the behaviour of the noise trader. The extent of the stock price variation in the passive case is small compared to the active case. Most importantly however, the fluctuations in the active case display a very clear recurring pattern while the variation in the passive case does not.

\begin{figure}
\centering
\includegraphics[width=0.7\textwidth]{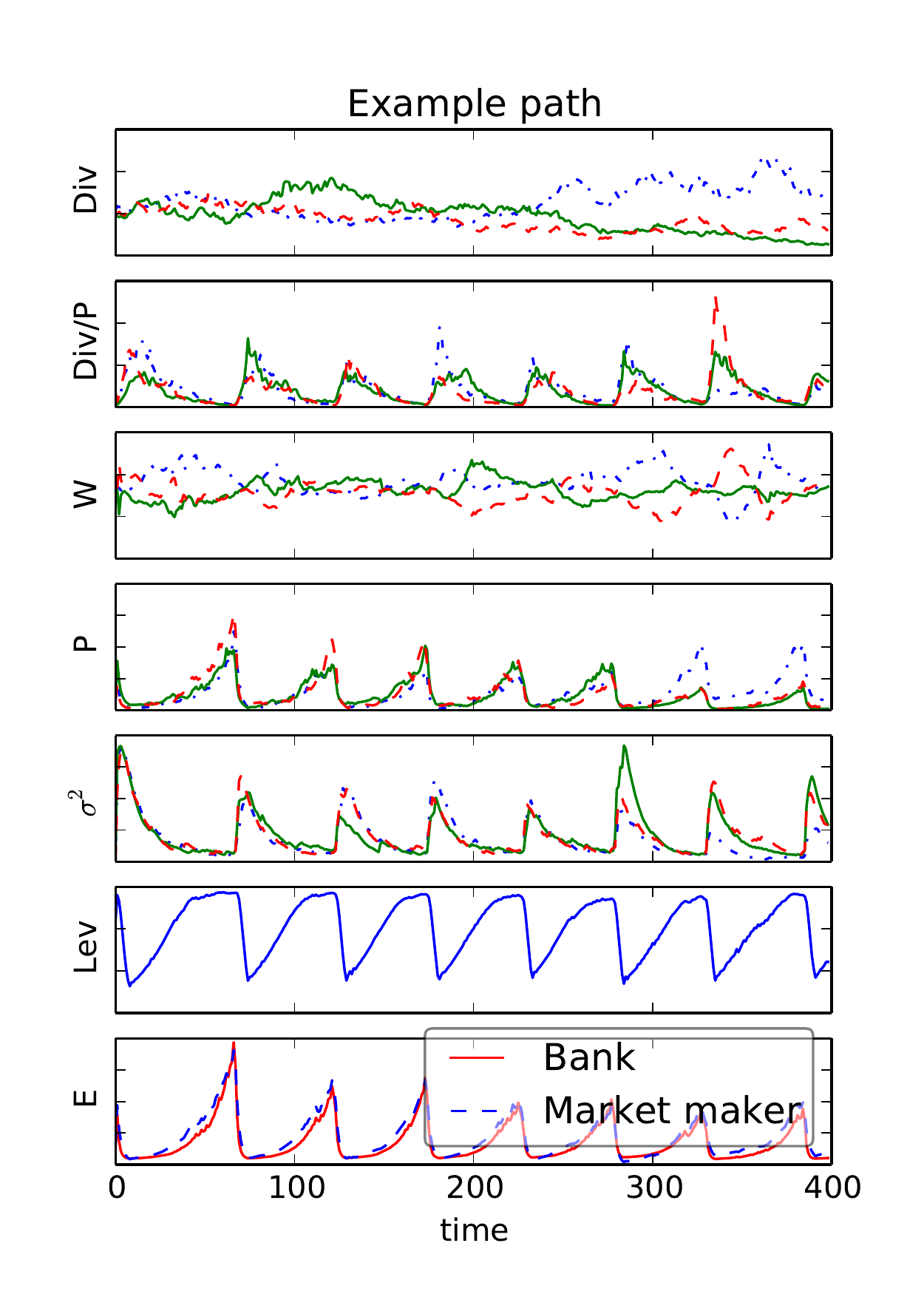}
\caption{Exemplary time series of the model with a bank that actively manages its leverage, a noise trader and three stocks. Time series from top to bottom: (1) stock dividends, (2) dividend price ratio, (3) bank portfolio weights,(4) stock prices, (5) leverage, (6) variance of stock prices, (7) equity of bank and noise trader. As mentioned before, the leverage management leads to recurring patterns of gradual price increases and drastic price crashes. Clearly the leverage is inversely related to the perceived portfolio risk and strongly positively related to the level of the stock market.   Price crashes occur typically when perceived risk is very low and leverage is very high, i.e. when banks perceive the world as very tranquil they are in fact at the highest risk. Note that leverage cycles affect all endogenous variables in the system, in particular leading to large variations in bank and noise trader equity.}
\label{FIG:full}
\end{figure}

In figure \ref{FIG:full} we expand our analysis to a wider set of model outputs.  As before, we are running the model with one bank, one noise trader and three stocks. We plot the following time series from top to bottom: (1) stock dividends, (2) dividend price ratio, (3) bank portfolio weights,(4) stock prices, (5) leverage, (6) variance of stock prices, (7) equity of bank and noise trader. 

A few points are worth noting:

Firstly, fluctuations in stock prices correlate strongly with fluctuations in leverage and fluctuations in leverage are anti-correlated to fluctuations in perceived portfolio risk. In fact, equation \ref{EQ::lev_target} implies that perceived portfolio risk drives the changes in leverage. Interestingly, perceived risk is lowest just before a crash, i.e. when the actual market risk is highest.

Secondly, stock prices are strongly cross correlated. While differences in the evolution of the stock's dividend process affect the stock prices, prices are modulated by the changes in leverage. This makes sense since changes in leverage determine the overall level of investment of the bank and thereby affect all stock prices.

Thirdly, fluctuations in stock prices affect the equity of all investors in the market, including the noise trader. Investors gain equity as the stock market rises and lose a substantial part of their wealth during a crash. While not shown here, in general these crashes can cause the bankruptcy of the investors. 

Finally, note that the variation in leverage, stock prices and equity significantly exceeds the underlying variation in the dividend process. This indicates that the leverage dynamics dominate the  effect of the portfolio rebalancing.  Taken together, figures \ref{FIG:active_passive} and \ref{FIG:full} strongly suggest that the recurring fluctuations observed in leverage, stock prices and equity result from the bank's leverage management.

We refer to the property of leverage management leading to the observed dynamics as pro-cyclical leverage. Furthermore, we refer to the fluctuations in leverage and stock prices as leverage cycles. This definition of pro-cyclical leverage extends the initial definition in \cite{Adrian2008a}. While \cite{Adrian2008a} refer to pro-cyclical leverage as the destabilizing effect of Value-at-Risk constraints in financial markets, figures \ref{FIG:active_passive} and \ref{FIG:full} illustrate that leverage management can lead to persistent fluctuations. Put in the language of dynamical systems, \cite{Adrian2008a} describe an unstable feedback loop. Here, we demonstrate that the dynamics induced by VaR constraints are richer and not necessarily fully unstable but potentially cyclical.

\section{Drivers of the leverage cycle}\label{Drivers}
In the previous section we demonstrated the existence of leverage cycles in a relatively rich agent-based model of a financial sector. The banks' behaviour is driven by two motives: adequately distributing their portfolio across assets based on their risk-return ratio and managing their risk by controlling their leverage. In order to better understand the drivers of the observed leverage cycles it is useful to consider simplified versions of the model. In particular we will consider two reduced models: (1) a two dimensional model with one asset and one investor with constant equity and (2) a five dimensional model with one asset, one investor and a noise trader, with variable equity.

The contribution of these reduced models will be two-fold. Firstly they allow a more analytical study of the system's dynamics and the development of an intuition for the drivers of the leverage cycle. Secondly, with the use of reduced models we can efficiently study the model dynamics in a large region of parameter space.

\subsection{Constant equity model}\label{2D}
In the following we reduce the agent-based model outlined in section \ref{MODEL} to a very simple two dimensional model, which has the advantage that it reproduces the essence of the leverage cycle and gives clear insight into its essential features.  To do this we assume a single asset and a single investor, which means that we can drop the indices $i$ and $j$ in equation \ref{EQ::stock_price}, and that the portfolio weight $w_{t+1} = 1$ and the fraction of the stock owned by the investor is $n_t = 1$.  The right hand side of equation (\ref{EQ::stock_price}) becomes $\mathcal{A}_{t}=  c_{t} + {p}_{t+1}$.  Furthermore, we assume that the equity of the investor is constant over time. While this is a strong assumption and a departure from the original model, it is underpinned by empirical research done by \cite{Adrian2008a}. In fact, \cite{Adrian2008a} show that investors respond to a change in assets by changing their leverage and keeping their equity fixed. We therefore consider it justified for the purposes of this section.

If the investor always maintains her target leverage then by definition $\overline{\lambda}(t) = \mathcal{A}_{t}/E$, where $E$ is the equity of the investor.  With these assumptions equation (\ref{EQ::stock_price})  reduces to $p_t = \mathcal{A}_{t}$, which can be written 
\begin{equation}
p(t) = \overline{\lambda}(t) E.
\end{equation}
As before, the estimated variance is given by equation (\ref{EQ::Cov_est}) and the target leverage  by equation (\ref{EQ::lev_target}),
\begin{equation}
\begin{aligned}
\sigma^2(t+1) &= (1 - \delta) \sigma^2(t) + \delta \left( \log \left( \frac{p(t)}{p(t-1)} \right) \right)^2, \\
\overline{\lambda}(t) &= \alpha \left( \sigma^2(t) + \sigma_0 \right)^b.
\end{aligned}
\end{equation}
In the following we will take $b = -0.5$ and $\sigma_0 = 0$. Eliminating prices from the system of equations we obtain
\begin{equation}
\sigma^2(t+1) = (1 - \delta) \sigma^2(t) + \frac{\delta}{4} \left( \log \left( \frac{ \sigma^2(t-1) }{\sigma^2(t)} \right) \right)^2. 
\end{equation}
We can re-write this as a two-dimensional deterministic dynamical system by writing $z_1(t) = \sigma^2(t)$ and $z_2(t) = \sigma^2(t-1)$, which gives
\begin{equation} \label{EQ::dynsys}
\begin{aligned}
z_1(t+1) &= (1 - \delta) z_1(t) + \frac{\delta}{4} \left( \log \left( \frac{ z_2(t) }{z_1(t)} \right) \right)^2, \\
z_2(t+1) &= z_1(t).
\end{aligned}
\end{equation}
In vector form this can be written $\mathbf{z}(t) = g(\mathbf{z}(t-1))$, where $\mathbf{z}(t) = (z_1(t) , z_2(t) )^T$.
While this simple model will be useful to understand the destabilizing effects of pro-cyclical leverage it has three important shortcomings:
\begin{itemize}
\item Firstly, the assumption that equity is fixed ignores an important aspect of leverage. Namely that if leverage is high, equity responds more drastically to shocks in assets. Therefore a bank with high leverage is inherently more risky than a bank with low leverage. This is not captured in this simple model.
\item  Secondly, the behavior of the model is independent of the riskiness of the bank $\alpha$. This means that the model is effectively insensitive to the level of leverage the bank takes.
\item Finally, the model is symmetric in the sign of the cyclicality parameter $b$. Therefore in this model we cannot distinguish between the effects of pro-cyclical and countercyclical leverage.
\end{itemize}
In order to address these shortcomings and study the dependence of pro-cyclical leverage on the parameters $\alpha$, $\delta$ and $b$ we will need to turn to a slightly more advanced model which will be introduced in section \ref{5D}.

\subsubsection{Intuition about system dynamics}
\begin{figure}
\centering
\includegraphics[width=0.8\textwidth]{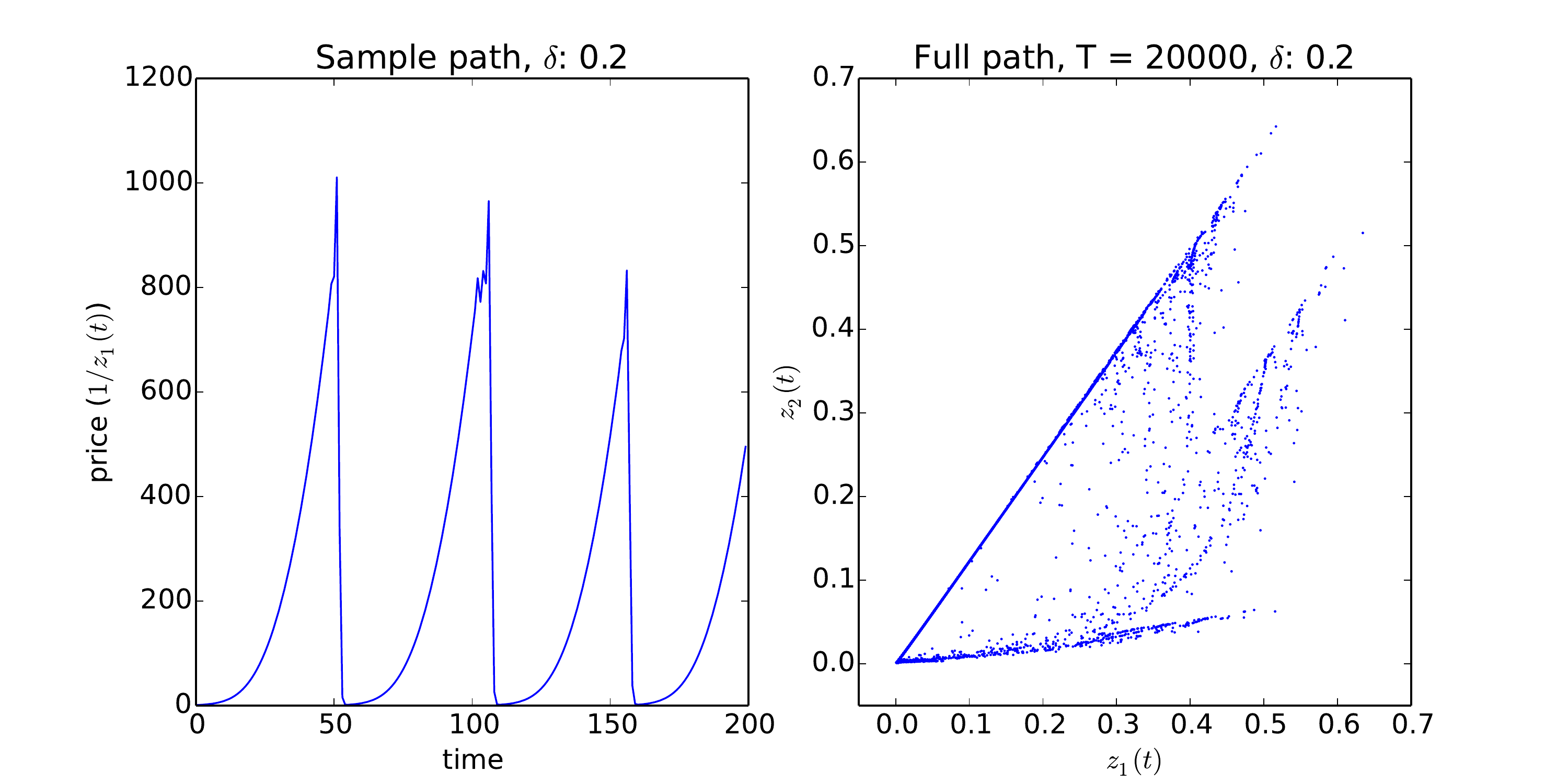}
\caption{Exemplary time series of the two dimensional model with one asset, one investor and constant equity. Left: We plot an extract of the price time series for 400 time steps. As in the full model, the simple model displays clear leverage cycles (recall that in the simple model prices are simply proportional to leverage). Right: We plot the evolution of the dynamical system for 20,000 time steps in a phase plot, i.e. on the x-axis we plot $z_1$ and on the y-axis we plot $z_2$. A state in the upper right corner of the figure moves along the straight nearly diagonal line towards the origin, then destabilises close to the origin and quickly returns to the straight line via a chaotic trajectory (often lingering briefly on the curve at the bottom of the figure.  The part of the trajectory on the straight line corresponds to the gradual increase in prices and leverage while the part of the trajectory below the straight line corresponds to the drastic price crash.}
\label{FIG:2D}
\end{figure}
We now develop an intuition for the dynamics of the system defined in equation (\ref{EQ::dynsys}). The left panel of figure \ref{FIG:2D} clearly illustrates that the basic features of the leverage cycle are preserved in this simple model: repeating patterns of gradual price increase followed by drastic price collapses. In the right panel of figure \ref{FIG:2D} we plot the evolution of the system spanned in the $z_1,z_2$ space for a long run with $T = 20,000$. The evolution of the system can be described follows:

Suppose we start in the top right corner of the phase space (the figure on the right), or equivalently at a trough of the time series plot of the price (the figure on the left).  This corresponds to the high volatility and low leverage regime, i.e. perceived risk is high ($z_1$) and was high in the previous time step ($z_2$).  In the phase space the system then moves along down the nearly  diagonal straight line towards the origin, corresponding to decreasing volatility and increasing leverage.  In the time series plot the price increases during this period, i.e. this is the bubble phase (e.g. the great moderation).  

At the origin the perceived risk is zero and the leverage is infinite.  Thus the system destabilizes as it approaches the origin, and a small perturbation is sufficient for the system to be ejected from its stable path towards the right of the phase plot where perceived risk is higher.  In the time series plot this corresponds to the initiation of a price crash.  The system then moves very quickly through the space below the diagonal of the phase plot until converging onto the high volatility and low leverage regime, where the price reaches a low value and the crash is over.   The leverage cycle restarts once the system has reached the high volatility and low leverage regime. The irregular nature of the time series and the fractal structure of the phase plot suggest that the system is not on a limit cycle, but rather a chaotic attractor (it is deterministic so these are the only two options).

From an economic perspective the approach towards the origin of the phase space along the straight line corresponds to the situation where perceived risk is low and continually decreasing. However, while perceived risk is low and decreasing, the system is actually moving towards a more and more unstable configuration in which small initial perturbations in perceived risk can cause rapid price movements and consequentially rapid adjustments of perceived risk.

The mismatch between perceived portfolio risk and actual systemic risk is crucial in the understanding of the dynamics observed here. As the bank learns about market risk it relies exclusively on historical data which falsely suggest an increasingly tranquil market. Instead, by adjusting to the lower perceived risk by taking on further leverage, the bank manoeuvres itself into regions of increased systemic risk.

\subsubsection{Linear stability analysis}
To make our analysis more formal, we will study the linear stability properties of the trajectory outlined above. Note that the system we are studying has a fixed point at $z^* = (0,0)$. This can be seen from equation (\ref{EQ::dynsys}). At a fixed point we must have $z_1 = z_2$ since $z_2$ is simply the previous position of $z_1$; at the fixed point $z_1$ is fixed hence $z_1 = z_2$. Taking the limit along $z_1=z_2$, we obtain $\lim_{(z_1,z_2) \rightarrow (0,0)} z_2/z_1 = 1$ due to l'Hospital's rule. Then the logarithm vanishes and $z_1$ is constant. Hence $z^* = (0,0)$ is a fixed point.

To compute the linear stability of the system we derive the Jacobian of the dynamical system and then study the value of the leading eigenvalue of the Jacobian in phase space. Recall the definition of the Jacobian: $J_{ij} = \partial g_i / \partial z_j$. Then:
\begin{equation}\label{EQ::Jacobian}
J = \left( \begin{matrix}
  1 - \delta - \delta \log\left( \frac{z_2}{z_1} \right) / (2 z_1) & \delta \log\left( \frac{z_2}{z_1} \right) / (2 z_2) \\
  1 & 0
 \end{matrix} \right).
\end{equation}
We can now diagonalise the Jacobian to obtain its eigenvalues. The absolute value of the largest eigenvalue of the Jacobian evaluated at the system's fixed point determines the stability of the system. Eigenvalues greater than 1 imply an unstable system while eigenvalues less than 1 imply a stable system.

We can now diagonalise the Jacobian to obtain its eigenvalues. We find for the eigenvalues:
\begin{equation}
\lambda_{\pm} = \frac{q_1(z_1,z_2,\delta) \mp q_2(z_1,z_2,\delta) }{q_3(z_1,z_2)}.
\end{equation}
For the exact functional form of the eigenvalues we refer the reader to appendix \ref{Eigenvalues2D}. When computing the eigenvalues at the fixed point $z^* = (0,0)$ we must be careful along which path to take the limit $(z_1,z_2) \rightarrow 0$. In general the limit will depend on the path taken. From figure \ref{FIG:2D} we note that the system approaches the origin on a straight line with slope $m>1$\footnote{It turns out that this slope is exactly $m = 1/(1-\delta)$.}. We therefore take the limit along the path $z_2 = m z_1$, where $m>1$. Then we obtain for the eigenvalues:
\begin{equation}
\begin{aligned}
\lim_{(z_1,m z_1)\rightarrow (0,0)} \lambda_- & = 1/m, \\
\lim_{(z_1,m z_1) \rightarrow (0,0)} \lambda_+ & = -\infty. \\
\end{aligned}
\end{equation}
The corresponding eigenvectors are: 
\begin{equation}
\mathbf{e}_{\pm} = (\lambda_{\pm},1)^T.
\end{equation}
Therefore $z^* = (0,0)$ is a hyperbolic fixed point which is stable along $\mathbf{e}_-$ and infinitely unstable along $\mathbf{e}_+$. The stable manifold corresponds to the nearly diagonal straight line $z_2 = mz_1$ along which we are descending towards the origin. The unstable manifold corresponds to $z_2 = 0$. 

As the system approaches the fixed point along the stable manifold it becomes increasingly susceptible to small perturbations. Once a small perturbation takes the system off the stable manifold it is ejected along the unstable manifold towards increasing $z_1$. Our stability analysis is therefore consistent with the computed trajectory of the system that we show in figure \ref{FIG:2D}.

The fact that the fixed point is hyperbolic is interesting from an economic point of view. For an agent that ``lives'' within this system, the dynamics may appear stable for long periods of time. The natural response for an investor in such a system is to take on more leverage. However, as noted above, it is precisely this response that moves the system closer to the hyperbolic fixed point and therefore makes it more susceptible to jumping to the unstable manifold. A system characterised by a hyperbolic fixed point can therefore transition very rapidly and unexpectedly from stable dynamics to highly unstable dynamics.

\subsection{Variable equity model}\label{5D}
In order to study the parameter dependence of pro-cyclical and countercyclical leverage as well as the impact of temporal control of the bank's riskiness $\alpha(t)$, we turn to a slightly more sophisticated model. In varying the parameters of the model developed in this section, we wish to identify regions of higher and lower systemic risk. Similarly, this model will allow us to test how effectively a regulator could control leverage cycles by varying bank riskiness over time. 

Rather than assuming constant equity at the outset, we compute the stock price from equation (\ref{EQ::stock_price}) for one bank (indexed by $B$) and one noise trader (indexed by $N$). Throughout this section we will assume that the stock investment of the bank is fixed at $w_B$ while the rest is invested into cash such that $w_c = 1-w_B$. For the noise trader we consider two regimes: (1) the stock investment weight follows a simple stochastic process or (2) the stock weight is fixed. Recall that $n(t)$ denotes the fraction of shares owned by the bank at time $t$. Further define the following terms:
\begin{equation}
\begin{aligned}
\overline{\lambda}(t) &= \alpha(t) \left( \sigma^2(t) + \sigma_0 \right)^b, \\
\Delta\mathcal{B}(t) &= \overline{\lambda}(t) (\mathcal{A}(t) - \mathcal{L}(t)) - \mathcal{A}(t), \\
\mathcal{A}(t) &= n(t)p(t)/w_B, \\
c_B(t) &= (1 - w_B)n(t)p(t)/w_B, \\
c_N(t) &= (1 - w_N(t))(1 - n(t))p(t)/w_N(t).
\end{aligned}
\end{equation}
$\overline{\lambda}(t)$ is simply the bank's target leverage as a function of the perceived portfolio risk. $\Delta\mathcal{B}(t)$ corresponds to the change in assets of the bank following a leverage adjustment. $\mathcal{A}(t)$ is simply the total value of the bank's assets prior to adding $\Delta\mathcal{B}(t)$. Similarly $\mathcal{L}(t)$ is the value of the bank's liabilities before adding $\Delta\mathcal{B}(t)$.  $c_B$ and $c_N$ are the cash held by the bank and the noise trader respectively. The bank's riskiness $\alpha(t)$ may change over time and can be controlled by a regulator. We will discuss the specification of the regulator's policy rule at the end of this section.
Given these definitions, the bank's perceived portfolio risk, liabilities, stock ownership as well as the stock price evolve as follows:
\begin{equation}\label{5D_prices}
\begin{aligned}
\sigma^2(t+1) &= (1-\delta) \sigma^2(t) + \delta \left( \log \left( \frac{p(t)}{p(t-1)} \right) \right)^2, \\
\mathcal{L}(t+1) &= \mathcal{L}(t) + \Delta\mathcal{B}(t), \\
n(t+1) &= \left( w_B(n(t)p(t+1) + c_B(t) + \Delta\mathcal{B}(t)) \right) / p(t+1), \\
p(t+1) &= \frac{w_B(c_B(t) + \Delta\mathcal{B}(t)) + w_N(t+1)c_N(t)}{(1 - w_Bn(t) - (1 - n(t))w_N(t))}.
\end{aligned}
\end{equation}
Note that for $\alpha(t) = const.$ this model specifies a five dimensional dynamical system with variables: $p(t), p(t-1), \sigma^2(t),\mathcal{L}(t), n(t)$. Further note that the expression for the perceived risk is just as before in the two dimensional dynamical system.

The bank's liabilities are updated simply based on how much the bank decided to change its balance sheet $\Delta\mathcal{B}(t)$. If $\Delta\mathcal{B}(t) > 0$ the bank borrows more and invests it into the stock market. If $\Delta\mathcal{B}(t) < 0$ the bank sells assets and pays back part of its debt.

The fraction $n(t+1)$ the bank will own of the stock depends on how much it invested into stocks and what the stock price turned out to be. Finally the stock price is proportional to how much both bank and noise trader want to invest into stocks. The denominator of the equation determining the stock price results from the fact that we are solving an implicit equation for the stock price, see equation (\ref{EQ::stock_price}). 

In the stochastic case the stock weight of the noise trader evolves as follows:
\begin{equation}\label{EQ::ns_w}
\begin{aligned}
\frac{dw_N(t+1)}{w_N(t)} &= (0.5-w_N(t))\rho + \eta dW, \\
w_N(t+1) &= w_N(t) + dw_N(t+1),
\end{aligned}
\end{equation}
where as before $\rho$ determines the speed of reversion to the portfolio balance. Note that we have removed the fundamentalist component of the noise trader's weight update since we are not considering the impact of dividends on the dynamics of this model. Making the stock weight of the noise trader stochastic adds one more dimension to the system and makes it six dimensional.

While in section \ref{2D} the bank's equity was fixed, here both the equity of the noise trader and of the bank can change according to the specification of the price process in equation (\ref{5D_prices}). In general, depending on the degree of variation in the noise trader's stock weight adjustment, either the bank or the noise trader lose money over the course of a simulation. In the short run, this is a positive feature of the model as it indicates the success of an investment strategy. In the long run however, it is problematic as eventually one investment strategy will dominate the market, and the dynamics are not stationary.  In order to impose stationarity we redistribute equity from the winning investor to the losing investor in every time step. In particular we adjust the above expressions for the bank's and noise trader's cash as follows:
 \begin{equation}
\begin{aligned}
c_B(t) &= (1 - w_B)n(t)p(t)/w_B +dE(t), \\
c_N(t) &= (1 - w_N(t))(1 - n(t))p(t)/w_N(t) -dE(t), \\
dE(t) &= \xi (E_0 - \mathcal{E}(t)),
\end{aligned}
\end{equation}
where $\xi$ is an adjustment rate, $E_0$ is the bank's equity target and $\mathcal{E}(t)$ is the bank's equity. Note that this approach ensures that equity in the system is conserved.  While being a convenient simplification in order to achieve stationary system dynamics and allow long run simulations, the stability properties of the system that we seek to investigate remain unaffected by this simplification, see Appendix \ref{EqDist}.

In the following we will propose a policy rule on the bank's riskiness $\alpha(t)$ imposed by a regulator who wishes to stabilize stock prices. Controlling $\alpha(t)$ is equivalent to adjusting the bank's required Value-at-Risk quantile over time.

Recall that the interaction between perceived risk and stock prices can lead to fluctuations in leverage and prices. An initial negative shock to stock prices increases perceived risk and tightens the bank's target leverage forcing. This forces the bank to reduce its investment which in turn leads to a further drop in price. Intuitively, in order to stabilize stock prices, the regulator should increase $\alpha(t)$ (increase the bank's leverage) when prices fall and decrease $\alpha(t)$ (decrease the bank's leverage) when prices rise. The bank's riskiness should therefore be inversely related to the return on stock prices. Note that this control policy is asymmetric in price changes while normal risk management policies are symmetric in price changes since risk is a function of squared returns. Given these initial considerations we propose the following policy rule for $\alpha(t)$:
\begin{equation}
\begin{aligned}
d \alpha(t+1) &= \rho_{\alpha} (\alpha_0 - \alpha(t)) + \theta q(t), \\
\alpha(t+1) &= \alpha(t) + d \alpha(t+1), \\
q(t+1) &= (1 - \delta_{\alpha}) q(t) + \delta_{\alpha} \log \left( \frac{p(t)}{p(t-1)} \right),
\end{aligned}
\end{equation}
where $\rho_{\alpha} < 1$, $\delta_{\alpha} < 1$ and $\theta$ are parameters. The regulator forms a temporal average of stock returns $q(t)$ with memory parameter $\delta_{\alpha}$. We refer to $q(t)$ as the stock price trend. If prices fall (negative returns), $q(t)$ will decrease and eventually become negative. Similarly, if prices recover (positive returns), $q(t)$ increases and becomes positive. This distinguishes $q(t)$ from the perceived risk $\sigma^2(t)$ which is measured in a very similar way but reacts symmetrically to increases and decreases in stock price.

The change in $\alpha(t)$ is then computed proportional to the stock price trend $q(t)$ scaled by the adjustment aggressiveness $\theta$. The larger $\theta$ the more the regulator will adjust $\alpha$ in response to a given price shock. Note that $\alpha(t)$ is anchored in a target value $\alpha_0$. This ensures that $\alpha(t)$ does not explode or go to zero. $\alpha(t)$ will revert back to $\alpha_0$ in the absence of price shocks. The idea is that the regulator uses small adjustments to stabilize the market rather than driving $\alpha$ to a specific value. Note that this policy rule includes two special cases:
\begin{enumerate}
\item $\theta = 0$: In this case the $\alpha(t) \rightarrow \alpha_0$. This is equivalent to the case where $\alpha(t)$ is pre-determined and fixed.
\item $\delta_{\alpha} = 1$: In this case the regulator reacts immediately to every shock, i.e. there is no smoothing of returns.
\end{enumerate}

\begin{figure}
        \centering
        \begin{subfigure}[b]{1.1\textwidth}
                \centering
                \includegraphics[width=\textwidth]{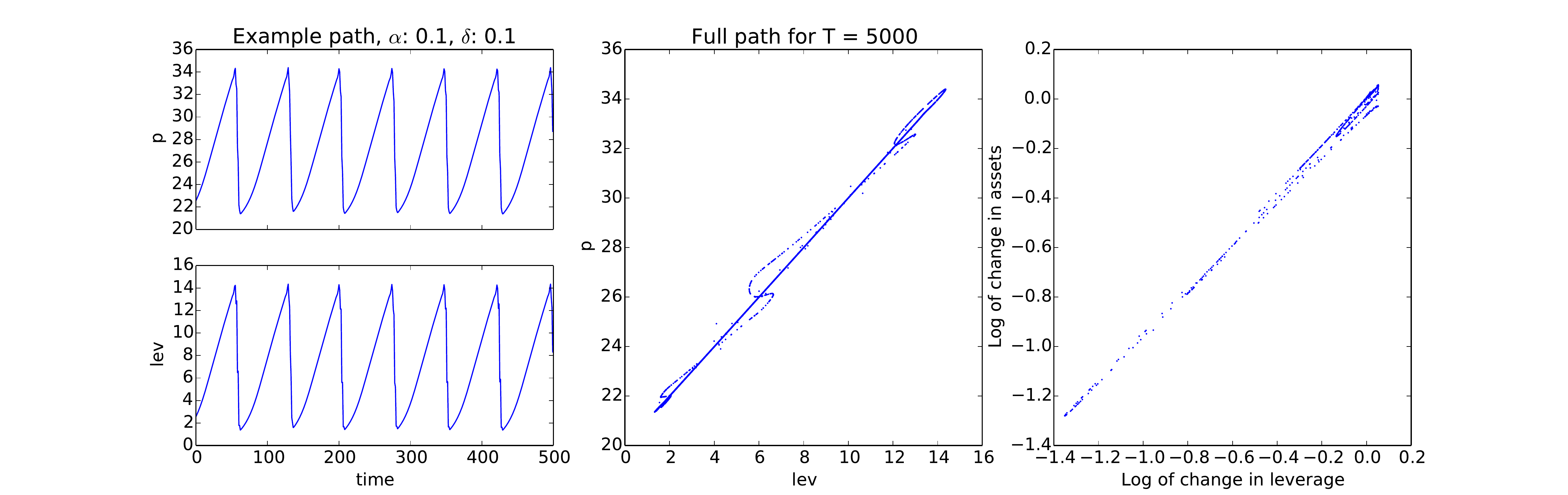}
        \end{subfigure}\\
        \begin{subfigure}[b]{1.1\textwidth}
                \centering
                \includegraphics[width=\textwidth]{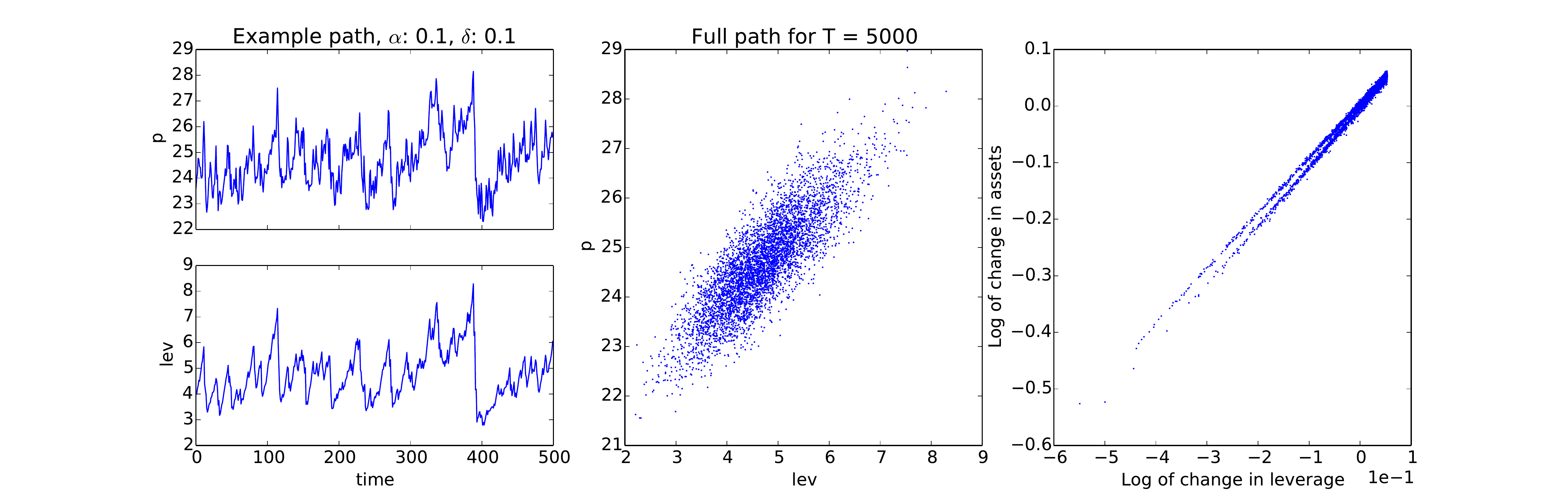}
        \end{subfigure}\\
         \caption{Example time series of reduced model with one bank, one noise trader and one stock. Parameters as indicated on plot and $b = -0.5$ and $\sigma_0 = 0$. Top panel: deterministic case with noise trader weights constant. Top left: time series for stock price and bank leverage for a 500 time step interval. Top middle: Phase plot of leverage vs stock price for 5000 time steps. Note the while the system dynamics in the short 500 time step interval resemble a limit cycle, the long run phase plot indicates that the system is in fact on a strange attractor. Top left: Scatter plot of log changes in assets vs log changes in leverage. This reproduces the positive correlation between changes in assets and changes in leverage originally noted in \cite{Adrian2008a}. Bottom panel: Stochastic case with noise trader weights evolving as specified in model. Bottom left: time series for stock price and bank leverage for a 500 time step interval. Bottom middle: Phase plot of leverage vs stock price for 5000 time steps. Bottom left: As for the deterministic case we show the positive correlation between log changes in assets and log changes in leverage. In both the stochastic and the deterministic case, the system clearly displays leverage cycles.}
         \label{FIG::5D_examp}
\end{figure}

We will use the simple variable equity model model outlined above to study four aspects of the bank's risk management strategy:
\begin{enumerate}
\item How do the stability properties of the system vary with the bank's riskiness $\alpha$ and its estimation horizon $\delta$?
\item What is the impact of counter-cyclical leverage policies (i.e. $b > 0$) on the stability of the system?
\item How can a policy rule on the bank's riskiness $\alpha$ stabilize the system?
\item To what extent can leverage limits stabilize the system?
\end{enumerate}
The parameters for all simulations presented in this paper are summarized in table \ref{TAB::param_overview_var_eq}. In all of the above except point $4$ we will have $\alpha$ fixed during a given simulation. Before we address each of the above points in sequence we discuss two exemplary time series generated by this model in figure \ref{FIG::5D_examp}:
\begin{itemize}
\item {\it Deterministic case}: We simulate the system for $T = 5000$ time steps with fixed weights for the noise trader. We plot an extract of this long run time series in the top left panel of figure \ref{FIG::5D_examp}. The dynamics closely resemble the dynamics observed in sections \ref{lev_cycles_m} and \ref{2D}. We observe recurring patterns of gradual price increases followed by rapid price crashes. Price movement correlates strongly with leverage adjustments. In the top middle panel of figure \ref{FIG::5D_examp} we plot the long run stock price series versus the leverage series in a so called phase plot. In this plot it becomes apparent that the system is not on a limit cycle. Instead its trajectory resembles a strange attractor with chaotic dynamics. Finally in the top left panel we show the positive correlation between changes in assets and changes in leverage that arises naturally in this model and is in accordance with empirical evidence, see \cite{Adrian2008a}:  The chaotic nature of the dynamical system is once again apparent in the structure of the point scatter.
\item {\it Stochastic case}: We simulate the system for $T = 5000$ time steps with stochastic weights for the noise trader as specified in equation (\ref{EQ::ns_w}). We plot an extract of this long run time series in the bottom left panel of figure \ref{FIG::5D_examp}. As in the deterministic case the system shows patterns of gradual price increases followed by rapid collapses. We plot the long run series in a phase plot in the bottom middle panel of figure \ref{FIG::5D_examp}. Here a clear positive correlation between stock prices and leverage is visible despite the stochastic component stemming from the noise trader. As in the deterministic case we plot in the bottom right panel, the changes in assets versus the changes in leverage. Again, we clearly observe the empirically documented positive correlation between changes in assets and changes in leverage.
\end{itemize}

\begin{table}
\begin{tabular}{lllr} 
\toprule
Agent & Parameter \\ 
\cmidrule{2-4}
& Notation & Description & Value \\
\midrule
Bank & $\delta$ & Memory parameter in covariance estimation & see run \\ 
	 & $\alpha$ & risk parameter & see run \\
	 & $b$ & cyclicality parameter & see run \\
	 & $E_0$ &  Bank equity target & $10$ \\ 
	 & $\lambda_{0}$ & Initial bank leverage & $5$ \\  
	 & $w_B$ & Bank stock weight & $0.05$ \\ 
	 & $n_0$ & Initial stock ownership & $0.1$ \\ 
	 & $\xi$ & Bank equity adjustment rate & $1.2$ \\	  
\midrule
Noise trader & $\rho$ & Portfolio balance parameter & $0.9$ \\ 	
			 & $\eta$ & Standard deviation Brownian motion & $0.01$ \\ 
\midrule
Policy rule & $\alpha_0$ & Equilibrium risk parameter & see run \\ 	
			& $\rho_{\alpha}$ & Relaxation rate & $0.5$ \\ 
			& $\delta_{\alpha}$ & Memory parameter for return estimation & see run \\
			& $\theta$ & Adjustment aggressiveness & see run \\  			
\bottomrule 
\end{tabular} 
\caption{Overview of simulation parameters for the variable equity model. In this section we sweep the first three parameters $\alpha$, $\delta$, $b$. The bank stock weight $w_B$ is simply $1-w_c$. $n_0$ sets the relative size of the bank to the noise trader. We chose our parameters such that the market impact of the bank is relatively low relative to the noise trader. We achieve this in particular by setting $w_B$ and $n_0$ to low values. Furthermore we choose the noise level from the noise trader to be comparatively low as we want to focus on the dynamics resulting from the leverage adjustment of the banks. The value for most parameters for the policy rule for the temporal control of the bank's riskiness depends on the particular run. We choose the relaxation rate to be relatively high in order to insure that the bank's riskiness returns quickly to the equilibrium value $\alpha_0$ that the regulator desires.}
\label{TAB::param_overview_var_eq}
\end{table}

\subsubsection{Riskiness and estimation horizon}
Intuitively lower bank riskiness and a longer estimation horizon should lead to a more stable system. As the riskiness $\alpha$ is decreased, the level of leverage decreases as well as the speed of leverage adjustment. Therefore decreasing $\alpha$ should always stabilize the system. The longer the estimation horizon (i.e. the smaller $\delta$) the slower the bank will respond to market shocks. A longer estimation horizon can be thought of introducing inertia that stabilizes the system. 

\begin{figure}
\centering
\includegraphics[width=0.45\textwidth]{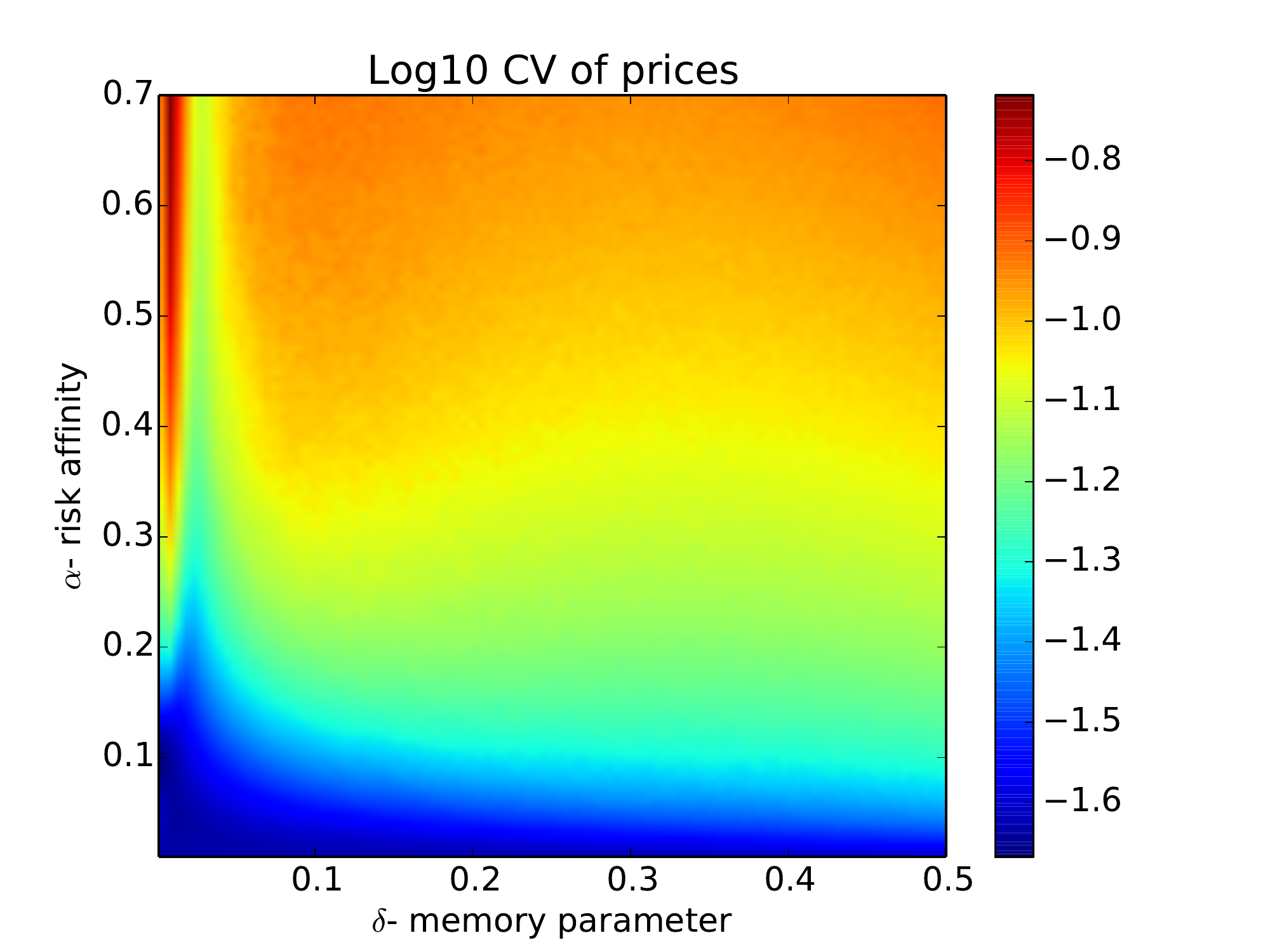}
\caption{Stability properties of the model as a function of bank riskiness and the risk estimation horizon. We plot $\log_{10}$ of the average coefficient of variation of the price time series for different values of bank riskiness $\alpha$ and estimation horizon $\delta$. Darker colors correspond to lower coefficient of variation. There is a region for low bank riskiness $\alpha$ and low $\delta$ (long estimation horizon) where the system stabilizes (blue region). Here the stochastic dynamics due to the noise trader dominate the dynamics induced by the bank's leverage management. As the bank riskiness is increased the variation of the prices increases across all values of $\delta$.}
\label{FIG:a_d_2_0}
\end{figure}

We investigate this intuition by running the model in the stochastic case for $T=5000$ with $\alpha \in \left( 0,0.7 \right]$, $\delta \in \left( 0,0.5 \right]$, $b = -0.5$ and $\sigma_0 = 0$.  For each run we record the coefficient of variation of the price time series ignoring the first $20 \%$ of the time series.  The coefficient of variation is defined as $CV = \sigma/\mu$, where $\sigma$ is the standard deviation and $\mu$ is the mean, and provides a useful nondimensional proxy for volatility. 
We then average over 40 seeds to obtain a value for the average coefficient of variation for each pair of $\alpha$ and $\delta$. We summarize our results in figure \ref{FIG:a_d_2_0}.

The space spanned by $\alpha$ and $\delta$ can be roughly divided into two regions. A relatively stable region (blue region) in which the coefficient of variation of the stock price is below $10^{-1.5}$ and a less stable region (green/orange region) above this threshold. In the stable region the dynamics due to the banks leverage management are dominated by the noise trader. Note that this region is restricted to relatively low values of bank riskiness ($\alpha \lesssim 0.15$) and thins out as the estimation horizon decreases ($\delta$ increases). Note that these values for $\alpha$ and corresponding leverage are indeed low considering that under the assumption of Gaussian returns and a VaR confidence level of $0.99$ we have $\alpha \approx 0.4$ and for the maximally heavy tailed distribution we have $\alpha \approx 0.17$.

Finally note that the size of the stable region actually decreases again for $\delta \lesssim 0.04$. We do not fully understand why we observe this cusp in the stable region but hypothesize that it is due to a characteristic time scale introduced by the mean reverting noise trader, which roughly matches.

\subsubsection{Counter-cyclical leverage policies}
Counter-cyclical leverage policies have been suggested in order to increase the stability of the financial system. The idea behind counter-cyclical leverage is to break the positive feedback loop outlined in \cite{Adrian2008a} and allow banks to take on more leverage as times get tough. Thus, a negative shock to stock prices can be followed by an increase in leverage rather than a drop in leverage as is the case in standard risk management approaches. Increased leverage pushes prices back up and thereby works against the initial price drop.

In this model the ``direction'' of leverage cyclicality is parametrized by the cyclicality parameter $b$. Thus we can study the impact of $b$ on the stability properties of the system on the continuum between fully pro-cyclical leverage ($b=-0.5$) and fully counter-cyclical leverage ($b=0.5$), corresponding to a simple inversion of the relationship between risk and leverage as implied by the Value-at-Risk approach.

In the following we study the coefficient of variation of the stock price in the deterministic case for $\alpha \in \left( 0,300 \right]$, $\delta = 0.1$, $b = \left[-0.5,0.5\right]$ and $\sigma_0 = 0$. In order to compute the coefficient of variation we run the simulation for $T = 5000$ and we ignore the first $20\%$ of the time series. For ease of exposition we plot the coefficient of variation only where it takes reasonable values. In the case of unstable behaviour we plot a default value of $10^5$. In case of bank bankruptcy we plot a default value of $10^4$. We mark a run as unstable if the price exceeds the maximum floating point number. Bankruptcy occurs simply when the equity of the bank falls below zero during the run.

\begin{figure}
        \centering
                \includegraphics[width=\textwidth]{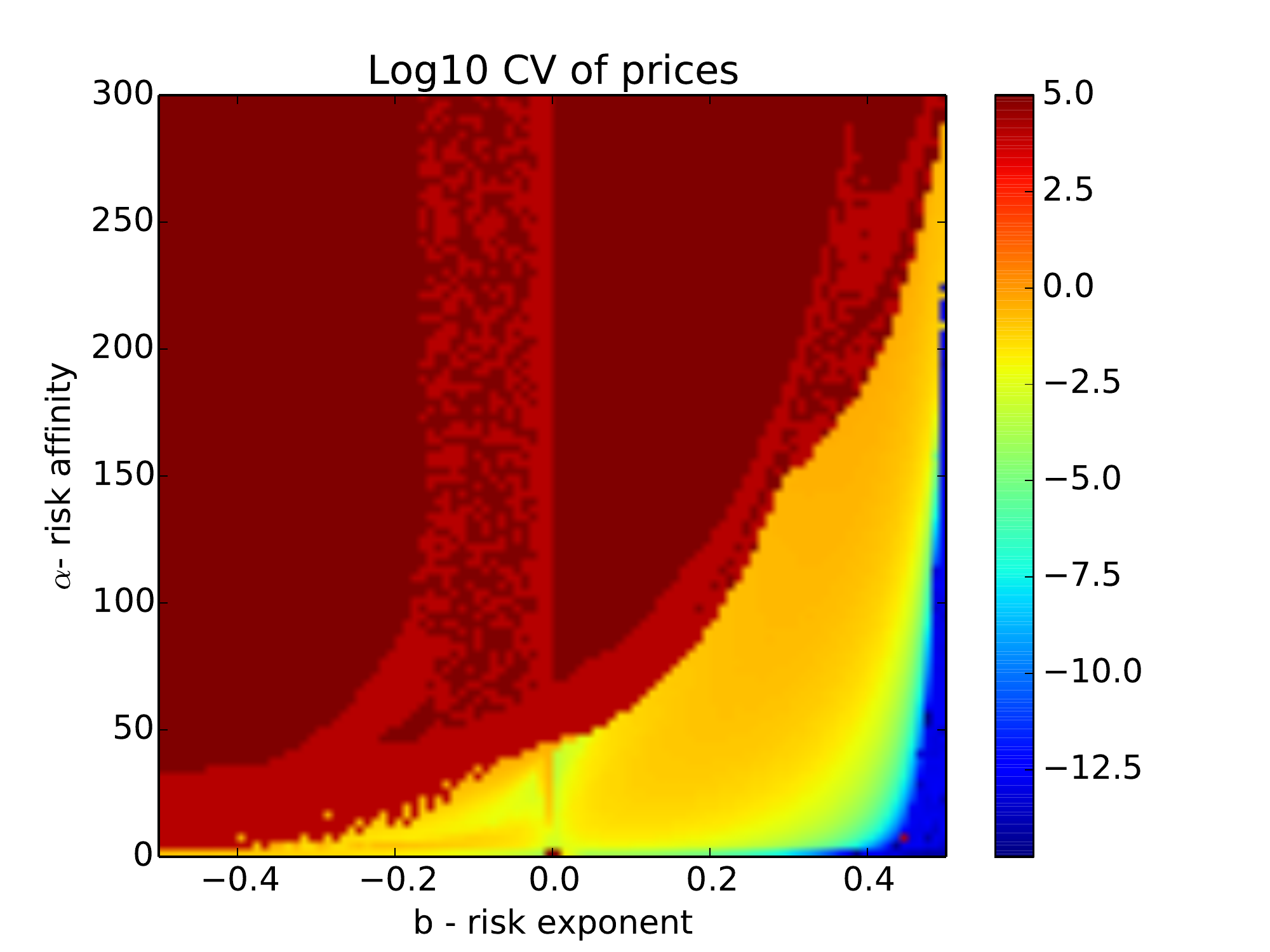}
         \caption{Stability properties of the model as a function of bank riskiness and the cylicality parameter $b$. We plot $\log_{10}$ of the coefficient of variation of the price time series for different values of bank riskiness $\alpha$ and cyclicality parameter $b$. Darker colors correspond to lower coefficient of variation. Dark and light red correspond to unstable dynamics. Left: full plot for $\alpha \in \left( 0,300 \right]$, $b = \left(-0.5,0.5\right]$. We plot a default value of $10^4$ for the coefficient of variation if the bank goes bankrupt during the simulation (light red) and a default value of $10^5$ (dark red) if the dynamics are unstable (i.e. a variable exceeds the maximum floating point number). Clearly the extent of the unstable region decreases as the cyclicality parameter $b$ increases from $-0.5$ to $0.5$. The yellow region displays leverage cycles while the blue region corresponds to a fixed point.}
         \label{FIG::a_b_5_0}
\end{figure} 

We summarize our results in figure \ref{FIG::a_b_5_0}. Visually we can identify roughly four different regions in the left plot in figure \ref{FIG::a_b_5_0}:
\begin{itemize}
\item Unstable region (dark red): For large values of bank riskiness $\alpha$ the system destabilizes. In this regime both leverage and prices explode. The size of this region decreases as the cyclicality parameter $b$ increases from $-0.5$ to $0.5$. As $b$ increases towards $0$ the bank adjusts its leverage downward less and less aggressively following a shock to perceived portfolio risk. Once $b$ becomes positive, shocks to perceived portfolio risk actually increase leverage rather than decreasing it. This behaviour counter-acts negative shocks to the stock price but amplifies positive shocks to price. Therefore counter-cyclical leverage does not eliminate instability entirely. 
\item Bankruptcy region (light red): At the lower boundary of the unstable region, banks tend to go bankrupt. In this region the leverage dynamics lead to a large surge in price followed by a sudden crash sufficiently large to wipe out the bank's equity.
\item Chaotic region (yellow): In this region of the parameter space the model displays fluctuations similar to the ones observed throughout this paper (see for example figure \ref{FIG::5D_examp}). The size of this region increases as $b$ increases. Importantly, the fluctuations persist and are of similar intensity for both $b < 0$ and $b > 0$. Thus, as already indicated by the existence of an unstable region for $b>0$, counter-cyclical leverage does not always dampen price deviations. This is due to the insensitivity of risk to the direction of price movement:  An increase in price leads to an increase in risk which leads to an increase in leverage which further increases the price. In this case the positive feedback loop runs in the opposite direction to the pro-cyclical case, i.e. positive price shocks are amplified whereas in the pro-cyclical case negative price shocks are amplified. Nonetheless, this feedback loop is damped by the noise trader and the countercyclical case is more stable than the pro-cyclical case. The stronger the pro-cyclicality, the higher the stability threshold.
\item Stable region (blue): For relatively low levels of bank riskiness $\alpha$ and large values of $b$ the system goes to a fixed point as is indicated by the vanishing coefficient of variation.
\end{itemize}
These findings show that risk management based on historical estimates of portfolio risk can give rise to a variety of dynamics ranging from stable to chaotic to unstable behaviour. The way banks react to changes in perceived risk and how much leverage they take is crucial in determining which behaviour the system will display.

\subsubsection{Temporal control of bank riskiness}\label{temp_con}
In the previous section we showed that a positive relationship between perceived risk and leverage can stabilize the system. However, it does not eliminate instability entirely since for $b>0$ positive shocks to the stock price are amplified. This behaviour is ultimately due to the symmetry of risk to changes in stock price: both positive and negative shocks can increase risk.

One way to break this symmetry is to apply a policy rule that is based on stock returns rather than squared returns as proposed above. In the following we study the coefficient of variation of the stock price under the policy rule specified above for $\alpha(t)$ with parameters $\delta_{\alpha} \in [0,0.7]$ and $\theta \in [0,7.0]$. In order to compute the coefficient of variation we run the simulation for $T = 5000$ and we ignore the first $20\%$ of the time series. Furthermore we take $\alpha_0 = 0.1$, $\delta = 0.1$, $b = -0.5$ and $\sigma_0 = 0$. We study the system in the deterministic case. 

\begin{figure}
        \centering
           \includegraphics[width=0.5\textwidth]{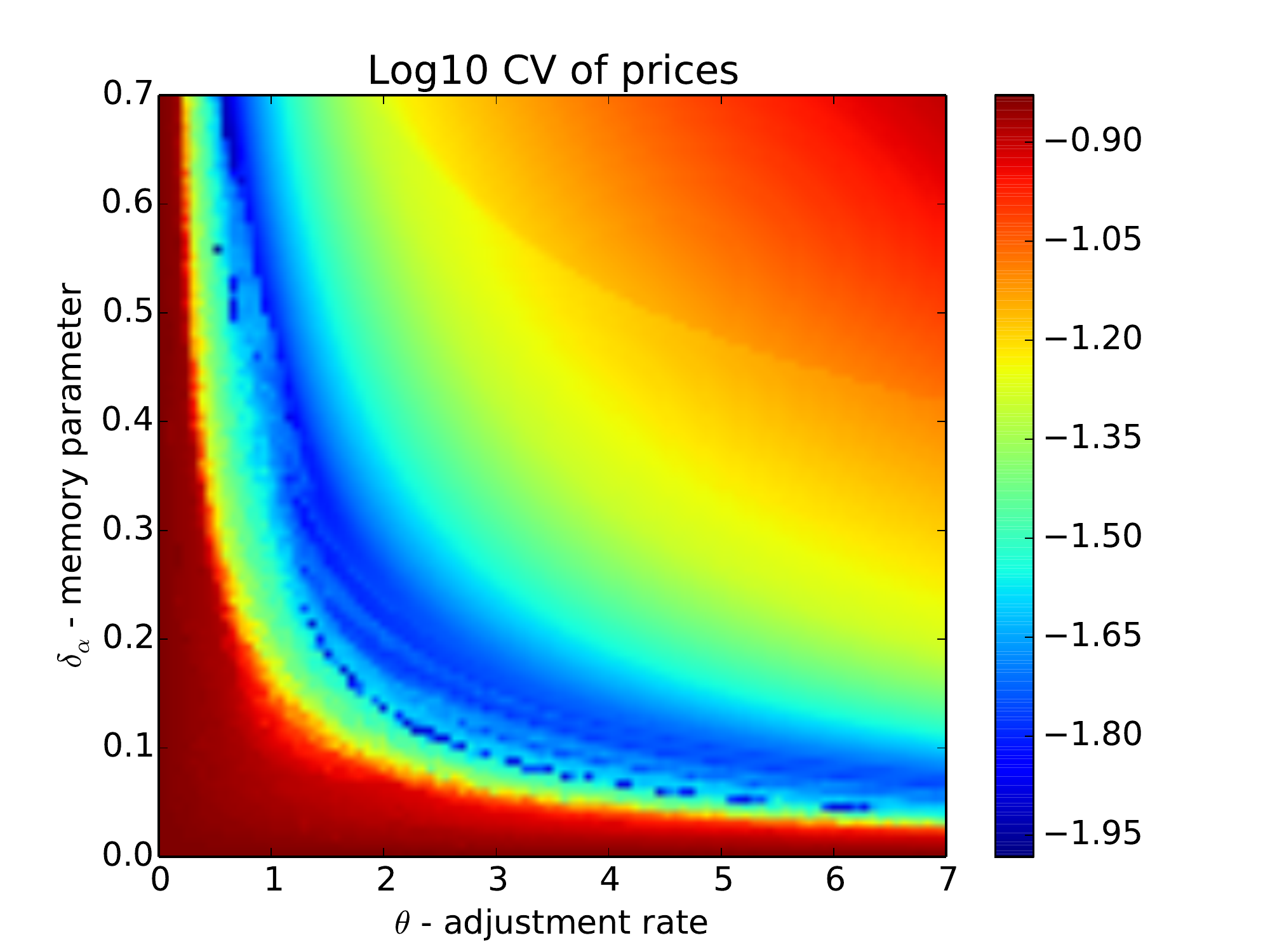}
         \caption{Stability properties of the model with temporal control of $\alpha(t)$ as a function of the adjustment aggressiveness and the estimation horizon of the stock price trend. We plot $\log_{10}$ of the coefficient of variation of the price time series for different values of the adjustment rate $\theta$ of $\alpha(t)$ and the memory parameter $\delta_{\alpha}$. For relatively low values of $\theta$ we recover the case where $\alpha$ is approximately fixed and observe comparatively high values for the coefficient of variation. Similarly for very low values of $\delta_{\alpha}$ the coefficient of variation is relatively high - the regulator does not respond quickly enough to price shocks. However, these relationships are non monotonic. As both $\delta_{\alpha}$ and $\theta$ are increased, first a region of relative stability emerges which then turns into a region of relative instability in the upper right quadrant. Here the regulator is over-reacting to changes in prices and destabilizes the system rather than stabilizing it.}
         \label{FIG::da_ata_0}
\end{figure} 

We summarize our results in figure \ref{FIG::da_ata_0}. Again, we can visually identify three regions:
\begin{itemize}
\item High CV region 1: For relatively low values of $\theta$ we recover the case where $\alpha$ is approximately fixed and observe comparatively high values for the coefficient of variation. Similarly for very low values of $\delta_{\alpha}$ the regulator does not respond quickly enough to price shocks and again $\alpha$ is approximately fixed. As for low $\theta$ the coefficient of variation of the stock price is relatively high in this region. Indeed, to some extent $\theta$ and $\delta_{\alpha}$ can compensate for each other producing the region to the right of the graph in which the CV is relatively high and roughly constant.
\item Low CV region: For intermediate values of $\theta$ and $\delta_{\alpha}$ the regulator successfully reduces the coefficient of variation. In fact the application of the temporal control of $\alpha$ reduces the CV by one order of magnitude. This is quite a substantial increase in stability.
\item High CV region 2: As $\theta$ and $\delta_{\alpha}$ is increased further, the CV increases again and returns to nearly as high values as in the region in the right of the plot. The non-monotonic behaviour is due to an overly strong forcing of the system by the actions of the regulator. The rule that was initially designed to stabilize the dynamics, destabilizes the system if applied to strongly.
\end{itemize}

Clearly, the application of a policy rule to control the bank's riskiness can help increase the stability of the system. Interestingly, there appears to be an optimal region in parameter space where the policy rule is most effective in stabilizing the system. This ``sweet spot'' corresponds to the blue band that is clearly visible in figure \ref{FIG::da_ata_0}. The rule is effective if the regulator's response is sufficiently forceful and quick to overcome the destabilizing effect of the bank's risk based leverage adjustment. However, the regulator's response should not be too aggressive. In fact, if she reacts too aggressively the level of the coefficient of variation returns to levels observed in the absence of the policy rule. This can be seen by comparing the coefficient of variation in the top right quadrant of figure \ref{FIG::da_ata_0} with the band on the very left where $\theta = 0$, i.e. the case without policy rule.

Increasing the level of $\alpha_0$ increases the overall level of the coefficient of variation, consistent with figures \ref{FIG::a_b_5_0} and \ref{FIG:a_d_2_0}, but does not change the qualitative structure of figure \ref{FIG::da_ata_0}. Similarly holding $\alpha_0$ fixed and increasing $b$ from $-0.5$ to $0.5$ decreases the overall level of the coefficient of variation while the qualitative structure remains unchanged at first. However, as seen in figure \ref{FIG::da_ata_0}, when $b$ becomes sufficiently large the system goes to a fixed point for sufficiently low $\alpha$. In this region, the impact of the policy rule vanishes since the system already goes to a fixed point.

\begin{figure}
        \centering
           \includegraphics[width=0.8\textwidth]{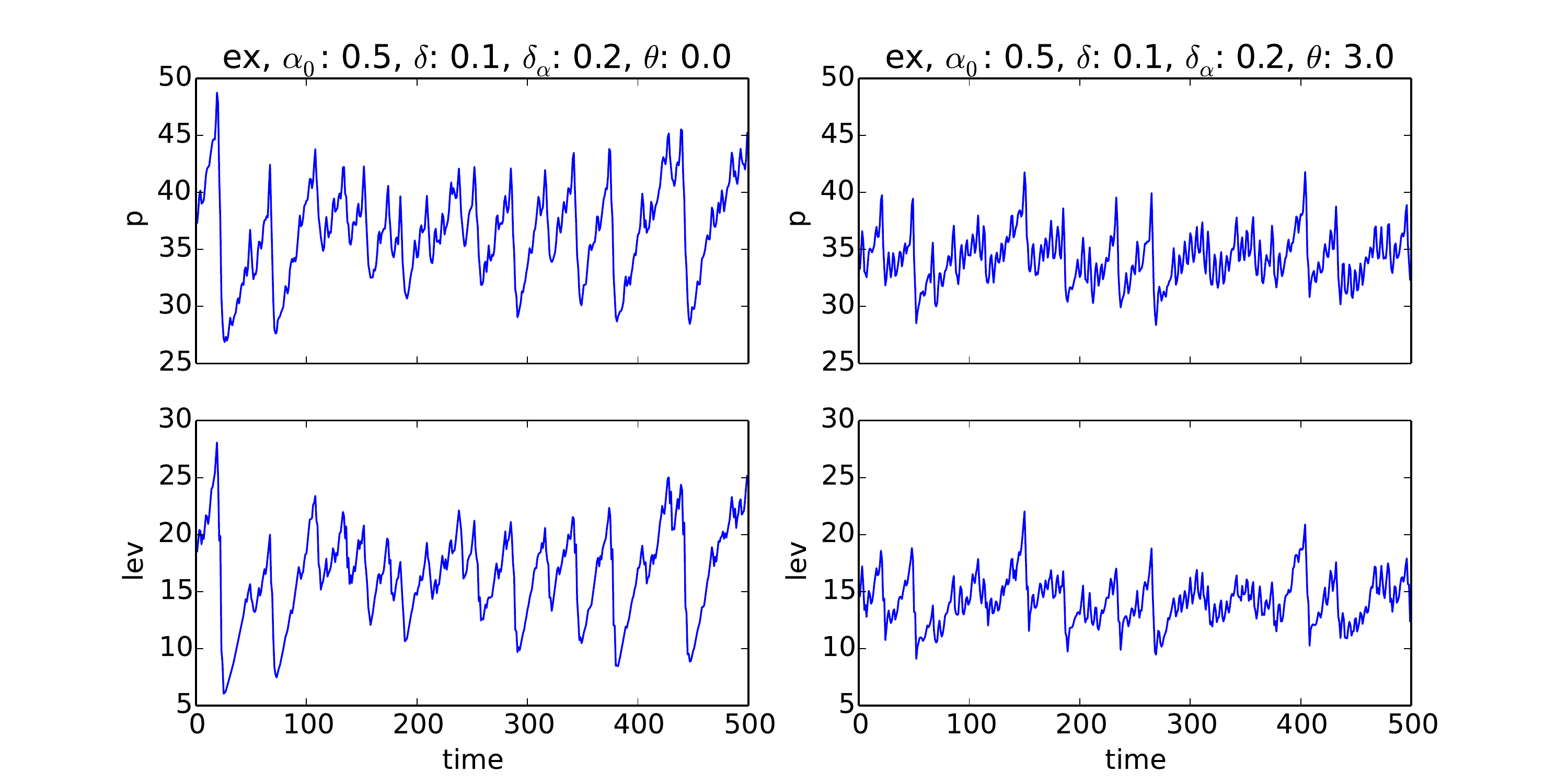}
         \caption{Price and leverage time series with noise for the cases with and without a policy rule on $\alpha$. Left panel: No policy rule. Right panel: With policy rule with parameters $\delta_{\alpha} = 0.2$ and $\theta = 3.0$. Note that the application of the policy rule maintains both average levels of price and leverage but reduces the amplitude of low frequency fluctuations. This comes at a cost of very high frequency fluctuations of low amplitude.}
         \label{FIG::combo_0}
\end{figure}

In order to further illustrate the impact of the policy rule on $\alpha(t)$ we present two time series in the stochastic regime: one with policy rule and one without. In particular we use the following parameters: $\alpha_0 = 0.5$, $\delta = 0.1$ and $\delta_{\alpha} = 0.2$. Recall that for $\theta = 0$ we recover the uncontrolled case. Therefore we use $\theta = 0$ for the case without policy rule and $\theta = 3.0$ for the case with policy rule. We present our results in figure \ref{FIG::combo_0} where we compare both stock and leverage time series for both cases.

Two aspects are worth noting. Firstly the application of the policy rule reduces the amplitude of the leverage cycle both measured by fluctuations in price and leverage. However, it does not achieve a complete elimination of the leverage cycles. Secondly, the reduction in the amplitude of low frequency leverage cycle comes at the cost of small amplitude high frequency oscillations that are introduced by the policy rule. The fact that the policy rule results in very fast changing values of $\alpha(t)$ raises questions about the practical applicability of such a rule. The frequency of these oscillations is likely to be inversely related to the parameter $\rho_{\alpha}$. 

\subsubsection{Leverage limits}
Another intuitive way of stabilizing the system would be via imposing leverage limits on banks. If the leverage limit is chosen sufficiently low, the leverage cycles shown throughout this paper should be damped if not eliminated entirely. We investigate this intuition by running our model in the stochastic case for a range of values for $\sigma_0$. Throughout this study we take $\alpha = 0.2$, $\delta = 0.1$ and $b = -0.5$. In order to give $\sigma_0$ a more intuitive interpretation we introduce the effective maximum leverage $\lambda_m = \alpha \sigma_0^b$. Note that in practice the maximum leverage imposed by a particular value of $\sigma_0$ will be lower than $\lambda_m$ since it ignores the actual perceived portfolio variance.

\begin{figure}
\centering
\includegraphics[width=0.8\textwidth]{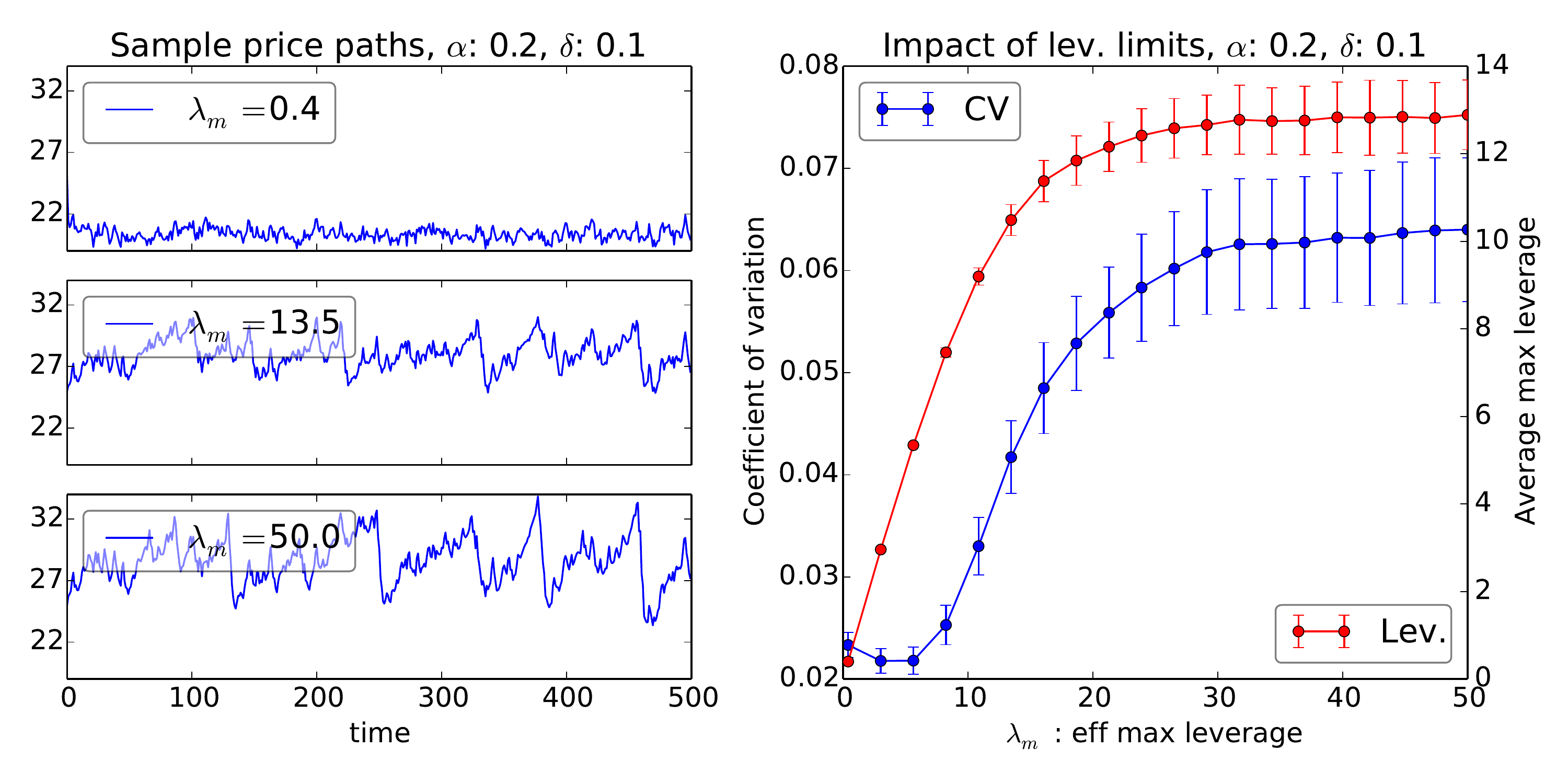}
\caption{Left: Three exemplary time series for different values of effective maximum leverage $\lambda_m$. Clearly the absolute level variation in the stock price time series increases as the leverage limit is increased. For a very low leverage limit the system effectively goes to a fixed point. Right: Right axis: average coefficient of variation of the stock price for different of $\lambda_m$. Left axis: Average maximum leverage of the bank for different $\lambda_m$. As the right plot already indicated, the coefficient of variation and absolute level of leverage increase as $\lambda_m$ is increased. However there is a $\lambda_m$ above which further increases in $\lambda_m$ have no effect in stabilizing the system. If the effective maximum leverage is above the bank's desired leverage level in the absence of a leverage limit, the leverage limit will have no effect. Note that the effective leverage limit $\lambda_m$ is not in one-to-one correspondence with the actual observed maximum leverage in the system. The observed maximum leverage is always less than or equal to $\lambda_m$ since the leverage target is computed not only on the basis of $\lambda_m$ but also the banks' perceived portfolio risk.}
\label{FIG:lev_lim_1}
\end{figure}

We summarize our results in figure \ref{FIG:lev_lim_1}. Clearly the variation in the price series and the maximum leverage of banks increases as we increase $\lambda_m$. The effect of $\lambda_m$ saturates as the system reaches its unconstrained target leverage. Since the bank will not increase its leverage beyond this value, an increased value of $\lambda_m$ will have no effect on the bank's leverage dynamics. Thus, not surprisingly, a maximum leverage will only be effective if chosen inside the actual decision interval of the bank.

\subsection{Policy implications}
In this paper we studied the stability implications of leverage management based on historical estimation of portfolio risk. From a policy perspective the insights resulting from this study have a number of implications:
\begin{itemize}
\item Risk appetite: The riskiness of a bank's leverage management strategy strongly affects the stability of the system. The riskier a bank's strategy the larger its impact on the leverage cycle. Its behaviour will lead to stronger price surges prior to a crash and more violent crashes. At the same time, once the bank's riskiness falls below a certain level, the system stabilizes. This somewhat discontinuous transition from stable to unstable behaviour is relevant for regulators as small changes to policy may have a large impact on the system wide dynamics.
\item Risk estimation horizon: Naively, fast updating of beliefs about market conditions should allow banks to better react to potential risks. While this may be true for an individual bank, the systemic impact of such behaviour is destabilizing. 
Our analysis indicates that shorter estimation horizons decrease the stability of the system up to a point:  Surprisingly, there is a finite time horizon that maximizes the stability of the system.
 \item Counter-cyclical leverage: Counter-cyclical leverage increases the system's resilience to negative price shocks and increases the ``safe operating zone" for the financial system.  It is not a panacea -- there is always a leverage ceiling above which the system is unstable.  Nonetheless, as we move from pro to counter-cyclical leverage the ceiling gets steadily higher.
 \item Temporal control of bank riskiness: The amplitude of leverage cycles can be reduced through the introduction of a stock return based policy rule on bank riskiness. However, two caveats apply. Firstly, a reduction in the amplitude of leverage cycles comes at the cost of introducing small amplitude - high frequency fluctuations. Secondly, a policy rule is only effective in a relatively small region of its parameter space. Careful calibration would therefore be required. 
\item Leverage limits: Leverage limits can be an effective tool to dampen leverage cycles if at set to an adequate level.
\end{itemize}
It is important to note that our results do not suggest that the absence of risk management is preferable over historically based risk management. Instead we take the existence of risk management as given for its obvious beneficial effect on the individual bank's balance sheet. Our focus is on how such risk management affects the stability of the system as a whole. The challenge for the regulator is then to identify regimes in which risk management is indeed stabilizing both on an individual level and on a systemic level.  This model is completely qualitative, but it nonetheless indicates the form that such tradeoffs may take.

\section{Conclusion}
In this paper we study the implications of historically based bank leverage management on the stability properties of the financial system. To this end we develop an agent-based model of a multi-asset financial system and study it computationally in its full form and two simplified versions. 

The banks' risk management is at the center of our model. In particular, we study banks that are subject to a Value-at-Risk constraint. In order to compute their Value-at-Risk, banks estimate the covariance matrix of their portfolio using an exponential moving average of the past returns of their investments. This set up establishes an inverse  relationship between perceived portfolio risk and the bank's desired leverage level. 

One central result of this paper is that the dynamics resulting from bank leverage management are richer than the unstable feedback between leverage, risk and asset prices (see \cite{Adrian2008a}) that is usually referred to a pro-cyclical leverage. In particular we show that bank leverage management can cause recurring patterns of stock price bubbles and crashes which occur in a chaotic regime of the system. We refer to these recurring bubbles and crashes as leverage cycles.

The intuition behind leverage cycles is simple: Suppose the bank's perceived portfolio risk is low and its leverage is high, i.e. the bank believes it is in tranquil times. Then an initial shock in the stock market leads to an increased perceived portfolio risk. When perceived risk is low and bank leverage is high, only a small shock to perceived risk will lead to a large adjustment of target leverage. This occurs simply by virtue of the functional form of the inverse relationship between risk and target leverage. A large adjustment in target leverage forces the bank to unload a large part of its assets. The associated market impact causes a further increase in perceived portfolio risk. Therefore a small initial shock can cause a downward spiral of prices and leverage. As leverage decreases more and more the bank's leverage adjustment rate to changes in market risk decreases. This property ultimately stabilizes the downward spiral and the market begins to recover. As the market recovers, perceived risk goes down and leverage and asset prices increase. This recovery sows the seeds for the next crash. 

The four main parameters of this model are: (1) the bank's riskiness $\alpha$, which determines the bank's level of leverage, (2) the portfolio risk estimation horizon $\delta$, (3) the cyclicality exponent $b$, which makes it possible to tune from pro-cyclical to countercyclical leverage  and (4) the effective maximum leverage $\lambda_m$. We vary these parameters to investigate and answer four main questions within this model:
\begin{itemize}
\item {\it How do the system's stability properties depend on the level of the banks' riskiness $\alpha$ and the smoothing parameter for volatility estimation, $\delta$}? Not surprisingly, increasing $\alpha$ tends to increase volatility and and if it is sufficiently large, it is always possible to destabilize the system completely.  (The only case where this is not true is when there is strong countercyclical leverage and $\alpha$ is sufficiently low that the system exists at a fixed point).  Varying $\delta$ gives more interesting results:  If $\delta$ is large, corresponding to a short estimation, decreasing $\delta$ (and thereby lengthening the horizon) reduces volatility and makes the system more stable.  Surprisingly, however, we find that there is a parameter $\delta_c$ where the stability reaches a maximum, and further decreasing $\delta$ make the system less stable.
\item {\it Under which circumstances do counter-cyclical leverage policies stabilize the system?}  Counter-cyclical leverage policies can stabilize the system if the cyclicality exponent is sufficiently large and the banks' riskiness is sufficiently small. This happens in two senses:  Starting from the purely pro-cyclical leverage ($b = -0.5$) and increasing $b$ to make the policy more countercyclical is clearly beneficial, raising the leverage ceiling where instability occurs.  For reasonably low risk levels, making the policy fully countercyclical ($b = 0.5$) can completely stabilize the system by driving it to a fixed point.  However, for large values of $\alpha$ counter-cyclical leverage policies still generate volatility and even instability. Therefore, while such policies do not solve the issue of leverage cycles in general, they can be effective in damping them.
\item {\it Can temporal control of the bank's riskiness $\alpha$ increase stability?} A regulator may decrease the amplitude of the leverage cycles by increasing $\alpha$ in response to negative price shocks and decreasing $\alpha$ in the case of positive price shocks. The effectiveness of this simple rule will depend on the time horizon over which the regulator measures the price movements and the aggressiveness of his response. Interestingly, this relationship is non-monotonic yielding a region of optimal measurement horizon and aggressiveness. However, control of leverage cycles comes at the cost of low amplitude - high frequency fluctuations.
\item {\it How effective are leverage limits ($\lambda_m$) in order to control leverage cycles?}  Leverage limits can curb leverage cycles effectively if set to an appropriate level. If the leverage limit exceeds the bank's target leverage in the absence of a leverage limit, the stock market dynamics remain unchanged by the introduction of the leverage limit. As the leverage limit is decreased, the amplitude of the leverage cycle decreases.
\end{itemize}

The model that is developed here is very simple and focuses on the impact of a single representative bank on the stock market behaviour. In the future it would be worthwhile to extend this model to a heterogeneous population of banks. Intuitively one would expect that heterogeneity in the portfolio positions of banks, their riskiness and their estimation horizon can increase the stability of the system. An interesting question would then be under which circumstances bank behaviour synchronizes. Then, under synchronized action we would expect to recover the dynamics observed for one representative bank.

A related extension would be to attempt a rough calibration of the model to a realistic financial system. A calibration would involve not only adjusting the relative sizes of the investors and their market impact but also the bank's response time to changes in leverage. We are currently assuming that the bank can adjust its leverage from one time step to the next. Introducing realistic constraints on this process is likely to increase the credibility of the model and allow the comparison with real financial time series.

\section{Acknowledgements}
This work was supported by the European Union Seventh Framework Programme FP7/2007-2013 under grant agreement CRISIS-ICT-2011-288501 and by Institute of New Economic Thinking at the Oxford Martin School. We would particularly like to thank Fabio Caccioli for very helpful discussions on the underlying mechanisms in our model, and laying the foundations with earlier work on a simple model.  Furthermore, we would like to thank Tobias Adrian and Nina Boyarchenko for their helpful insights provided during the CRISIS workshop in Leiden and via email. We would also like to thank three anonymous reviewers for their very helpful comments. Finally we would like to thank Olaf Bochmann for his efforts in developing the CRISIS software and useful feedback on the coding implementation of this model.

\bibliography{Citations.bib} 
\bibliographystyle{chicago}

\newpage
\appendix
 
\section{Eigenvalues of constant equity model}\label{Eigenvalues2D}
In the following we will derive the eigenvalues of the constant equity model in section \ref{2D}. Recall the Jacobian of the dynamical system:
\begin{equation}
J = \left( \begin{matrix}
  1 - \delta - \delta \log\left( \frac{z_2}{z_1} \right) / (2 z_1) & \delta \log\left( \frac{z_2}{z_1} \right) / (2 z_2) \\
  1 & 0
 \end{matrix} \right).
\end{equation}
We can now diagonalise the Jacobian to obtain its eigenvalues. We find for the eigenvalues:
\begin{equation}
\lambda_{\pm} = \frac{q_1(z_1,z_2,\delta) \mp q_2(z_1,z_2,\delta) }{q_3(z_1,z_2)},
\end{equation}
where
\begin{equation}
\begin{aligned}
q_1(z_1,z_2,\delta) &= -2 \delta z_1 z_2-\delta z_2 \log \left(\frac{z_2}{z_1}\right)+2 z_1 z_2,\\
q_2(z_1,z_2,\delta) &= \sqrt{8 \delta z_1^2 z_2 \log \left(\frac{z_2}{z_1}\right)+\left(2 \delta z_1 z_2+\delta z_2 \log \left(\frac{z_2}{z_1}\right)-2 z_1 z_2\right)^2}, \\
q_3(z_1,z_2) &= 4 z_1 z_2.
\end{aligned}
\end{equation}
The corresponding eigenvectors are: 
\begin{equation}
\mathbf{e}_{\pm} = (\lambda_{\pm},1)^T.
\end{equation}

\section{Alternative portfolio allocation\label{AppendixPortfolio}}
In the main model we propose a simple portfolio allocation rule based on the Sharpe ratio of a particular stock. In the following we will discuss an alternative portfolio allocation approach based on portfolio optimization subject to a Value-at-Risk constraint. The bank chooses the vector of portfolio weights $\mathbf{w}$ in order to maximize:
\begin{equation}
\begin{aligned}
& \underset{\mathbf{w}}{\text{max}}
& & \mathbf{w} \cdot \hat{\mathbf{r}} \\
& \text{subject to}
& & \mathcal{E} - a \mathbf{w}^T \mathbf{\Sigma} \mathbf{w} \geq 0,\\
& & & w_i \geq 0, \forall i
\end{aligned}
\end{equation}
where $\hat{\mathbf{r}}$ is the vector of expected stock returns, $\mathcal{E}$ is the bank's equity and $\mathbf{\Sigma}$ is the bank's estimated portfolio variance. This problem cannot be solved analytically due to the no-shorting constraints on the portfolio weights. In the implementation of the simulation we use a numerical optimizer to compute the results. 

\begin{figure}
\centering
\includegraphics[width=0.8\textwidth]{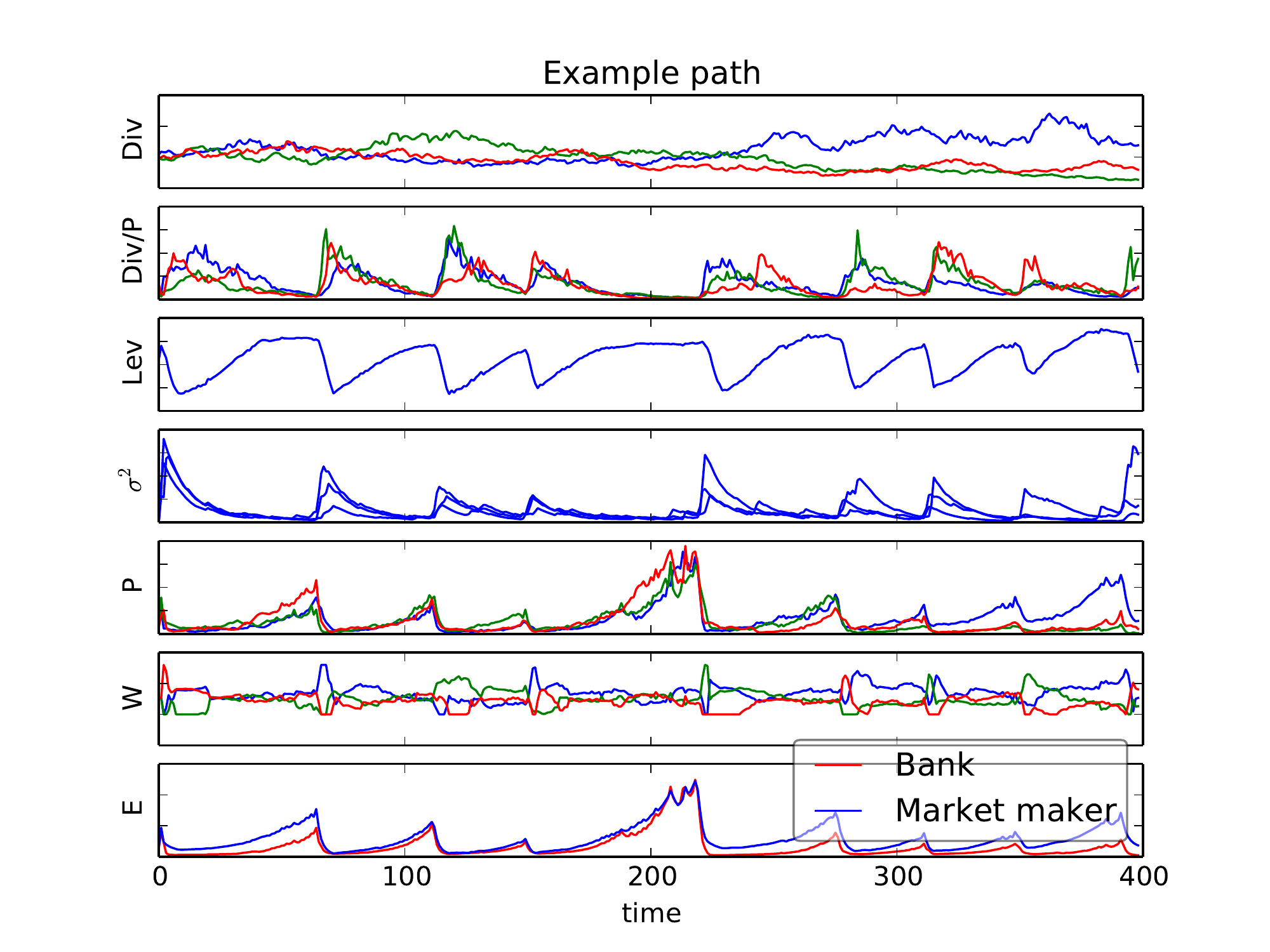}
\caption{Exemplary time series of the model with a bank that actively manages its leverage. Time series from top to bottom: (1) dividends, (2) dividend price ratio, (3) leverage, (4) variance of stock prices, (5) stock prices, (6) bank portfolio weights, (7) equity of bank and noise trader.}
\label{FIG:alt_port}
\end{figure}

In figure \ref{FIG:alt_port} we show an example run for a simulation with all parameters as in table \ref{TAB::param_overview} in section \ref{sim_Set_up}. However instead of using the simple portfolio allocation based on the Sharpe ratio we solve the above optimization problem to compute the portfolio weights. The main dynamics, namely leverage cycles, remain unchanged under this alternative portfolio allocation scheme. Since our focus is on leverage cycles, we chose the simpler portfolio allocation method for computational efficiency. The simple portfolio allocation method runs approximately 20 times faster than the method based on numerical optimization.

\section{Effect of equity redistribution on model dynamics}\label{EqDist}
In section \ref{5D} we introduce a redistribution mechanism of equity from the noise trader to the bank and vice versa. We introduce this mechanism in order to make the model dynamics stationary. In the absence of the equity redistribution the noise trader gradually accumulates equity since his passive trading strategy is superior to the bank's active strategy in the presence of leverage cycles. This non-stationarity makes the interpretation of long simulation runs difficult. Thus the equity redistribution mechanism is a very useful feature. 

\begin{figure}
\centering
\includegraphics[width=0.5\textwidth]{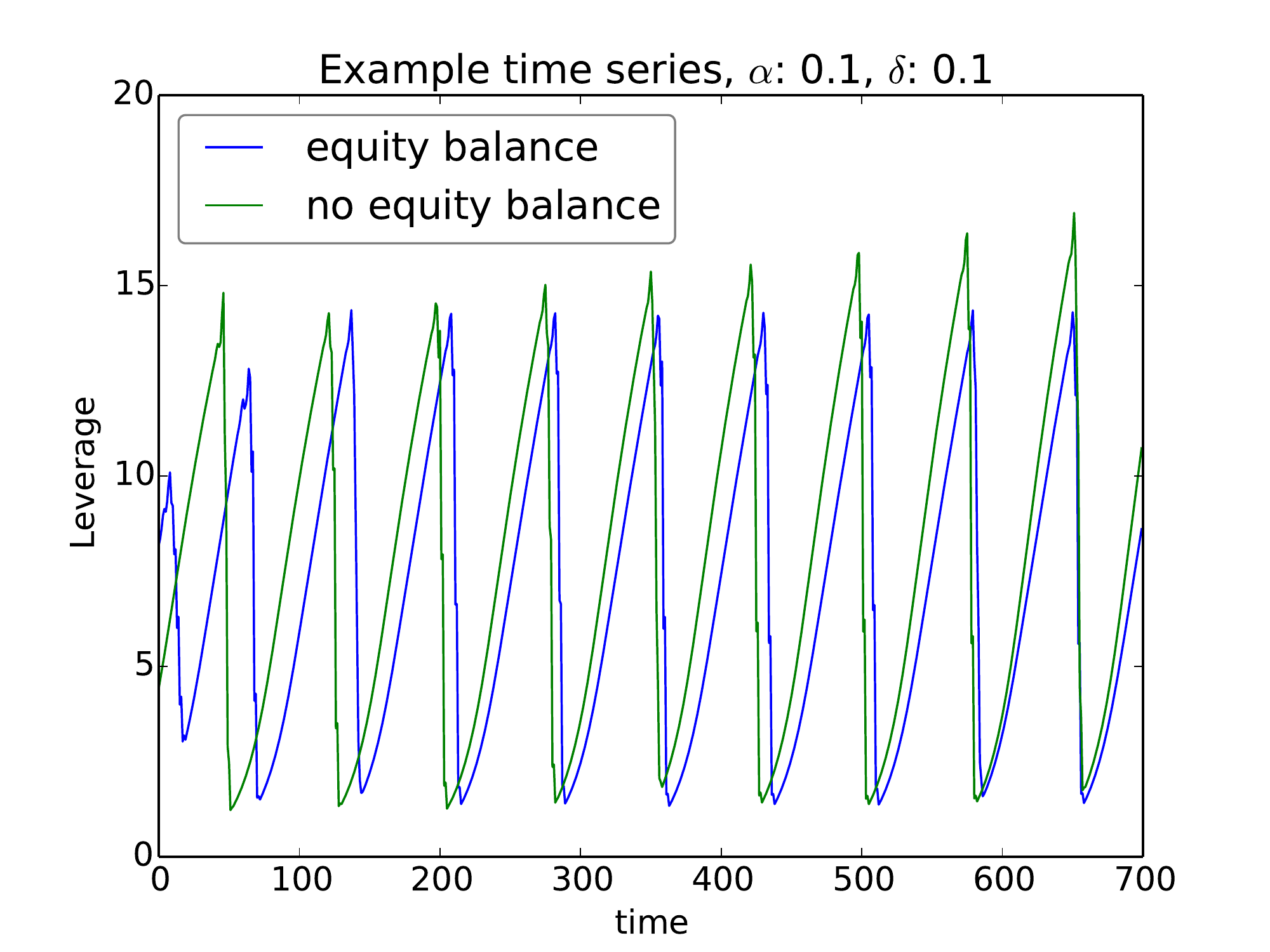}
\caption{Exemplary leverage time series of the model presented in section \ref{5D} for standard parameters as indicated on the plot and summarized in table \ref{TAB::param_overview_var_eq}. The blue time series corresponds to the case with equity redistribution while the green time series corresponds to the case without equity redistribution. Clearly both time series display leverage cycles of similar frequency and amplitude. However, the cycle in the absence of equity redistribution slowly increases in amplitude. This is due to the fact that the equity of the bank slowly declines over the cycles. This non-stationarity is removed by the equity redistribution mechanism.}
\label{FIG:eq_dist}
\end{figure}

In the following we illustrate that the equity redistribution mechanism does not affect the principal dynamics of the model presented in section \ref{5D}. In figure \ref{FIG:eq_dist} we show two exemplary leverage time series of the model presented in section \ref{5D} for standard parameters as indicated on the plot and summarized in table \ref{TAB::param_overview_var_eq}. The blue time series corresponds to the case with equity redistribution while the green time series corresponds to the case without equity redistribution. Clearly both time series display leverage cycles of similar frequency and amplitude. However, the cycle in the absence of equity redistribution slowly increases in amplitude. This is due to the fact that the equity of the bank slowly declines over the cycles. This non-stationarity is removed by the equity redistribution mechanism.

\end{document}